# Tuning the magnetic properties in MPS$_3$ (M = Mn, Fe, and Ni) by proximity-induced Dzyaloshinskii Moriya interactions

Suvodeep Paul*$^a$, Devesh Negi$^a$, Saswata Talukdar$^a$, Saheb Karak$^a$, Shalini Badola$^a$, Bommareddy Poojitha$^a$, Manasi Mandal$^a$, Sourav Marik$^a$, R. P. Singh$^a$, Nashra Pistawala$^b$, Luminita Harnagea$^b$, Aksa Thomas$^c$, Ajay Soni$^c$, Subhro Bhattacharjee$^d$, and Surajit Saha*$^a$

a. Department of Physics, Indian Institute of Science Education and Research Bhopal, Madhya Pradesh, India

b. Department of Physics, Indian Institute of Science Education and Research Pune, Maharashtra, India

c. School of Physical Sciences, Indian Institute of Technology Mandi, Himachal Pradesh, India

d. International Centre for Theoretical Sciences, Tata Institute of Fundamental Research, Bangalore, India

* Corresponding authors: suvodeeppaul100@gmail.com, surajit@iiserb.ac.in

## Abstract

Tailoring the quantum many-body interactions in layered materials through appropriate heterostructure engineering can result in emergent properties that are absent in the constituent materials thus promising potential future applications. In this article, we have demonstrated controlling the otherwise robust magnetic properties of transition metal phosphorus trisulphides (Mn/Fe/NiPS$_3$) in their heterostructures with Weyl semimetallic MoTe$_2$ which can be attributed to the Dzyaloshinskii Moriya (DM) interactions at the interface of the two different layered materials. While the DM interaction is known to scale with the strength of the spin-orbit coupling (SOC), we also demonstrate here that the effect of DM interaction strongly varies with the spin orientation/dimensionality of the magnetic layer and the low-energy electronic density of state of the spin-orbit coupled layer. The observations are further supported by a series of experiments on heterostructures with a variety of substrates/underlayers hosting variable SOC and electronic density of states.

Controlling the quantum many-body interactions is the key to discovering emergent phenomena and exploring potential technological applications. This may be achieved by the influence of various stimuli like the effect of reduced dimensionality[1], the application of an electric or a magnetic field[2], ultrafast laser pulses[3], proximity to suitable substrates *etc.*[4,5]. The proximity effect, in two-dimensional (2D) layered materials, may be induced by creating appropriate heterostructures. Engineered heterostructures of 2D materials is one of the ways to effectively control the quantum many-body interactions. The 2D layered magnets exhibit novel magnetic properties wherein heterostructure engineering could introduce effects of broken inversion symmetry as well as spin-orbit and Dzyaloshinskii Moriya (DM) interactions, thus resulting in new exotic ground states. Examples of such ground states are topologically protected spin textures like skyrmions and chiral domain walls[6–10] which can have novel spin-orbitronic and storage applications[11,12].

Transition metal phosphorus trisulphides (MPS$_3$, M = Mn, Fe, and Ni) are a class of such van der Waals materials that host antiferromagnetic (AFM) ground states at low temperatures[13]. The AFM ground state exhibits different spin dimensionalities, (*viz.*, $n = 1, 2,$ and $3$) due to the presence of an axial or planar anisotropy or in the absence of any anisotropic element[14] which may be described by the Ising (*e.g.*, FePS$_3$), XY (*e.g.*, NiPS$_3$), and Heisenberg (*e.g.*, MnPS$_3$) Hamiltonians, respectively[13]. Engineered heterostructures of magnetic layered materials with high spin-orbit coupled systems like topological materials has the potential to control the quantum interactions unravelling a variety of exotic phenomena at their interfaces[4,15,16]. We have performed a comprehensive study on the heterostructures of various MPS$_3$ compounds (M = Mn, Fe, and Ni) with the topological Weyl semimetal T$_d$-MoTe$_2$ that serve as an excellent platform for exploring the rich interfacial phenomena like Rashba effect, DM interactions, and the effect of spin orientations and dimensionalities on their properties.

Raman spectroscopy is a non-destructive characterizing tool for low-dimensional crystals that can provide information about the sample's structure, phonon properties, couplings between the various degrees of freedom, and many-body interactions involving low-energy excitations[17–19]. The magnetic ordering in a material affects the phonon properties due to spin-phonon interactions[3,20] or magnetoelastic couplings, and these signatures can be observed through Raman scattering. Therefore, Raman measurements are very well suited for the detection of spin ordering in low dimensional magnets[21–23], especially, for micron-sized flakes and heterostructures where standard bulk measurement techniques like SQUID-VSM are not very suitable.

Here, we bring out the robustness[21–23] of the magnetic ordering in MPS$_3$ compounds (M = Mn, Fe, and Ni) to flake thickness. More importantly, our measurements on MPS$_3$/MoTe$_2$ heterostructures demonstrate a strong suppression of the magnetic ordering for MnPS$_3$ and FePS$_3$ while negligible effects on NiPS$_3$, which could have its origin in the spin, orbit, and charge couplings at the interface mediated by DM interactions. The interfacial interactions reveal strong sensitivity to the orientation of the spins in the magnetic layer and the low-energy electronic density of states of the non-magnetic layer (substrate/underlayer). Our observations are further supported by experiments on heterostructures of MPS$_3$ with Sb$_2$Te$_3$, Au, and Cu with varying SOC and density of states.

## Results

This section begins with a brief description of the crystal structure and spin orientations corresponding to the various investigated MPS$_3$ magnets along with the signatures of the magnetic transition as observed through Raman measurements. This will be followed by the effects of dimensionality and proximity to the high SOC underlayers of topological Weyl semimetal T$_d$-MoTe$_2$, topological insulator Sb$_2$Te$_3$, as well as metallic Au and Cu.

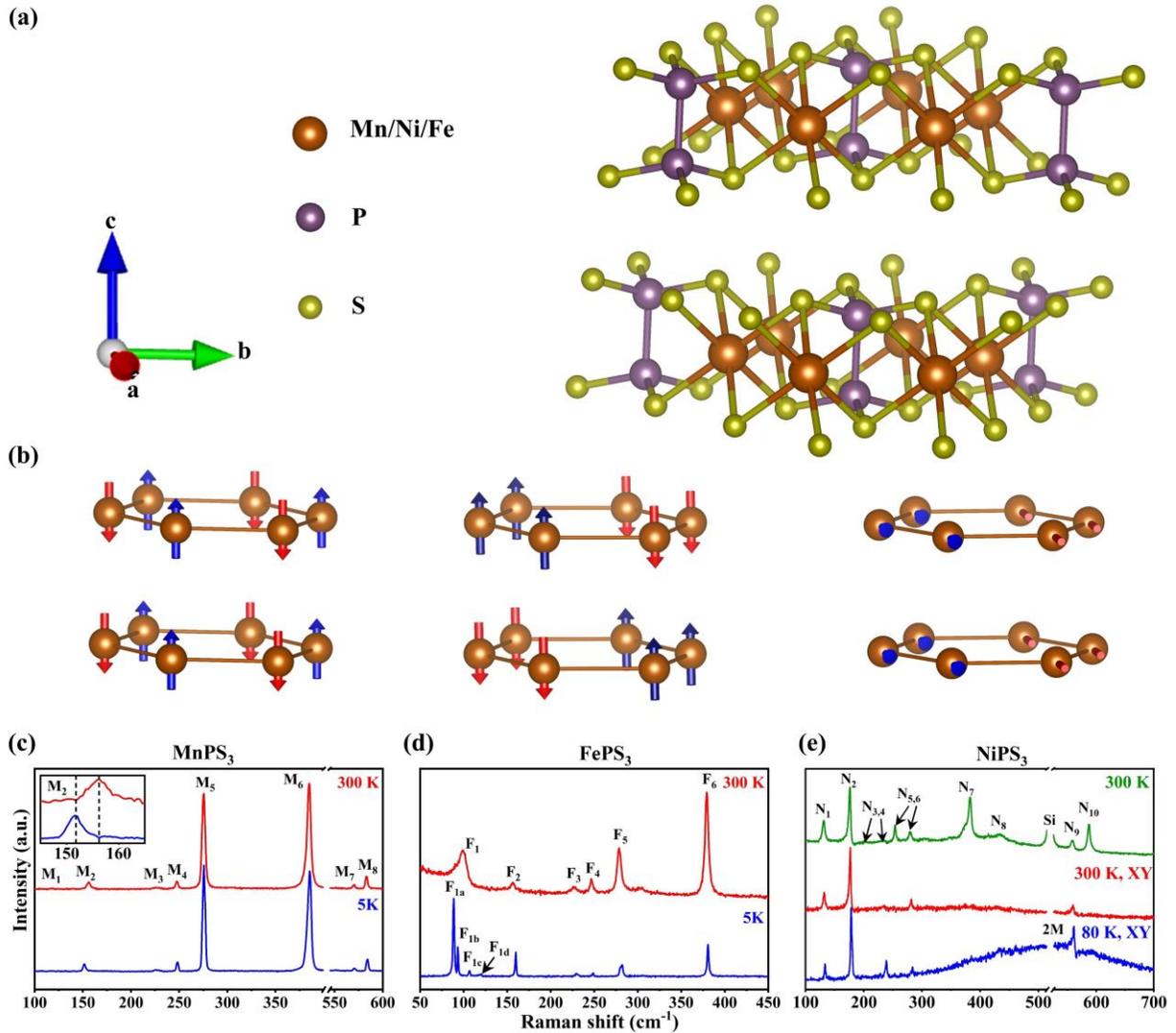

**Figure 1. (a)** The crystal structure of MPS$_3$ (M= Mn, Fe, and Fe), **(b)** The spin structure of MnPS$_3$, FePS$_3$, and NiPS$_3$ in their corresponding AFM phases, **(c)** A comparison between the Raman spectra of MnPS$_3$ in their PM (obtained at 300 K) and AFM (obtained at 5 K) phases. (Inset) An enlarged view of the M$_2$ phonon which shows the strongest signature of the transition, **(d)** A comparison between the Raman spectra of FePS$_3$ in their PM (obtained at 300 K) and AFM (obtained at 5 K) phases, **(e)** The room temperature (300 K) Raman spectrum of NiPS$_3$ (green). A comparison between the cross-polarized Raman spectra of NiPS$_3$ in their PM (obtained at 300 K) and AFM (obtained at 80 K) phases.

**Crystal and spin structures and their Raman signatures**

The MPS$_3$ compounds possess a monoclinic crystal structure, as shown in Fig. 1a, with a space group of C2/m[24,25]. The transition metal M atoms form a honeycomb structure where each M atom is coordinated to six S atoms in a trigonal symmetry, thereby, forming an MS$_6$ octahedron with its trigonal axis perpendicular to the basal plane of the crystal. The S atoms are further bonded to two P atoms above and below the M plane arranged in the shape of a dumbbell.

The MPS$_3$ compounds undergo a transition from the paramagnetic (PM) to the AFM phase at the corresponding Neel temperatures (T$_N$'s) which are reported as 80 K, 115 K, and 150 K for MnPS$_3$, FePS$_3$, and NiPS$_3$, respectively[13]. The magnetization data are shown in Figure S1 of Supplementary information. The spin structures associated with the AFM phases of these compounds have been extensively studied through neutron diffraction experiments[26,27]. Though the MPS$_3$ compounds are isostructural, they exhibit contrasting spin structures in their respective AFM phases, as shown in Fig. 1b. It has been reported that a trigonal distortion of the MS$_6$ octahedra results in different spin structures in these compounds[13]. Both MnPS$_3$ and FePS$_3$ have their magnetic moments oriented out-of-plane (parallel to the trigonal axis of the MS$_6$ octahedra) and the AFM orderings occur at wave vectors $\boldsymbol{k} = [000]$ and $\boldsymbol{k} = \left[00\frac{1}{2}\right]$, respectively[26]. However, more recent studies have reported the magnetic propagation vector for FePS$_3$ as $\boldsymbol{k} = \left[01\frac{1}{2}\right]$[28,29]. NiPS$_3$, on the other hand, has all the magnetic moments oriented in-plane (perpendicular to the trigonal axis of the MS$_6$ octahedra) along the a-axis and adopts a wave vector of $\boldsymbol{k} = [010]$ in its AFM phase[27]. Furthermore, while MnPS$_3$ shows antiferromagnetic ordering with respect to the nearest neighbors in the honeycomb lattice formed by the Mn$^{2+}$ ions, FePS$_3$ and NiPS$_3$ gives the appearance of two antiferromagnetically ordered chains of ferromagnetically ordered Fe$^{2+}$/Ni$^{2+}$ ions[24,26,27,30]. However, unlike MnPS$_3$, FePS$_3$ exhibits unidentical magnetic and crystallographic unit cells[26].

While the Raman signatures of all the MPS$_3$ compounds are similar in their PM phase on account of their structural equivalence, we observe drastically different signatures of the magnetic transition as their magnetic structures and the wavevectors associated with the AFM orderings are different. The room temperature (300 K) Raman spectra for MnPS$_3$, FePS$_3$, and NiPS$_3$ are shown in Figure 1c-e and the corresponding Raman modes are labelled as M$_1$-M$_8$, F$_1$-F$_6$, and N$_1$-N$_{10}$, respectively. The modes M$_1$ and M$_2$ (for MnPS$_3$), F$_1$ and F$_2$ (for FePS$_3$), and N$_1$ and N$_2$ (for NiPS$_3$) are associated with vibrations of the respective transition metal ions (*i.e.*, Mn$^{2+}$, Fe$^{2+}$, and Ni$^{2+}$), while the other modes with higher energies correspond to the vibrations of the (P$_2$S$_6$)$^{4-}$ units. The room temperature Raman spectra of MPS$_3$ (in their PM phase) are compared with the corresponding spectra in their AFM phase (obtained at 5 K for MnPS$_3$ and FePS$_3$, and 80 K for NiPS$_3$), as shown in Fig. 1c-e, to bring out the distinct Raman signatures of their AFM ordering.

Interestingly, all the phonon modes (M$_1$-M$_8$) of MnPS$_3$ observed at room temperature, are also present in its AFM phase (Fig. 1c). The phonon M$_2$ (and also M$_1$ which is very weak because of its low Raman cross-section) shows a large anomalous redshift in the low temperature spectra, as observed in the enlarged view of the $M_2$ phonon in the inset of Fig. 1c. The observed redshift is in contradiction to the expected blueshift at low temperatures due to anharmonicity. The anomaly is a result of the magnetoelastic coupling associated with the $M_2$ phonon[31]. Therefore, the evolution of the $M_2$ phonon as a function of temperature has been established and extensively used as the signature of the transition of $MnPS_3$ from its PM to the AFM phase[23,31].

In the case of $FePS_3$ (Fig. 1d), we observe the appearance of certain new modes at low temperatures in addition to the modes (labelled as $F_2$-$F_6$) observed at room temperature. The $F_1$ mode (associated with the vibration of $Fe^{2+}$ ions), which is broad and asymmetric at room temperature, splits into four sharp peaks ($F_{1a}$, $F_{1b}$, $F_{1c}$, and $F_{1d}$) and shows a dramatic enhancement in the intensity below the $T_N$[21]. Additionally, the $F_2$ phonon (also associated with the vibration of $Fe^{2+}$ ions) shows an appreciable enhancement in intensity below the $T_N$[21]. These effects can be attributed to the phenomenon of zone-folding due to the doubling of the magnetic lattice with respect to its crystallographic unit cell induced by the magnetic ordering in $FePS_3$. As a result, Raman inactive zone boundary phonons are converted to Raman active zone center phonons, giving rise to new Raman modes in the AFM phase[21]. The mode $F_2$ is associated with a Γ-point phonon and therefore, is visible even above the $T_N$ (unfolded Brillouin zone). But the local fluctuations and disorders in the crystal reduce its intensity above the ordering temperature. Below $T_N$, the long-range spin ordering is expected to overcome the effects of local distortions and hence sharp intense peaks are observed[21].

$NiPS_3$ (Fig. 1e), on the other hand, shows ten Raman active modes labelled as $N_1$-$N_{10}$ at low temperature (80 K). In order to better resolve the differences between the spectra obtained in the PM and AFM phases, we have compared the spectra obtained in the perpendicular polarization configuration, where the low temperature spectrum clearly exhibits a broad peak near ~550 cm$^{-1}$, which is attributed to the two-magnons (2M) scattering[22]. The 2M feature enhances very rapidly below the $T_N$ (refer to Supplementary information), thus signifying the AFM ordering in $NiPS_3$. Furthermore, the $N_2$ mode (near ~178 cm$^{-1}$) constitutes of two nearly degenerate components, such that one component is observed in the xx (parallel) polarization configuration, while the other is observed in the xy (perpendicular) configuration. As discussed in Supplementary information, the difference in frequencies between these constituent peaks vanishes above $T_N$. Therefore, the difference in frequencies of these constituent components of $N_2$ has also been considered as an order parameter for the detection of the AFM state in $NiPS_3$, particularly in multilayered $NiPS_3$[22]. However, in case of monolayer flakes, due to the lack of monoclinic stacking and the averaging effect of multiple magnetic domains, this criterion fails, and therefore, gives misleading information about the $T_N$ [30]. Furthermore, the $N_2$ mode (in both polarized and unpolarized configurations) shows a clear departure from the anharmonic behavior below $T_N$ due to spin-phonon coupling. The onset of the spin-phonon coupling can be used to identify the AFM spin ordering temperature[32].

**Effect of dimensionality and proximity on the AFM ordering**

We have conducted extensive studies on a series of samples of MnPS$_3$, FePS$_3$, and NiPS$_3$ of varying thicknesses using various underlayers/substrates. A description of the investigated samples is provided in the Supplementary information (See Note SN2 and Figure S2). The corresponding thicknesses of the samples have been determined through atomic force microscopy measurements shown in Supplementary information (Note SN3 and Figures S3-S9).

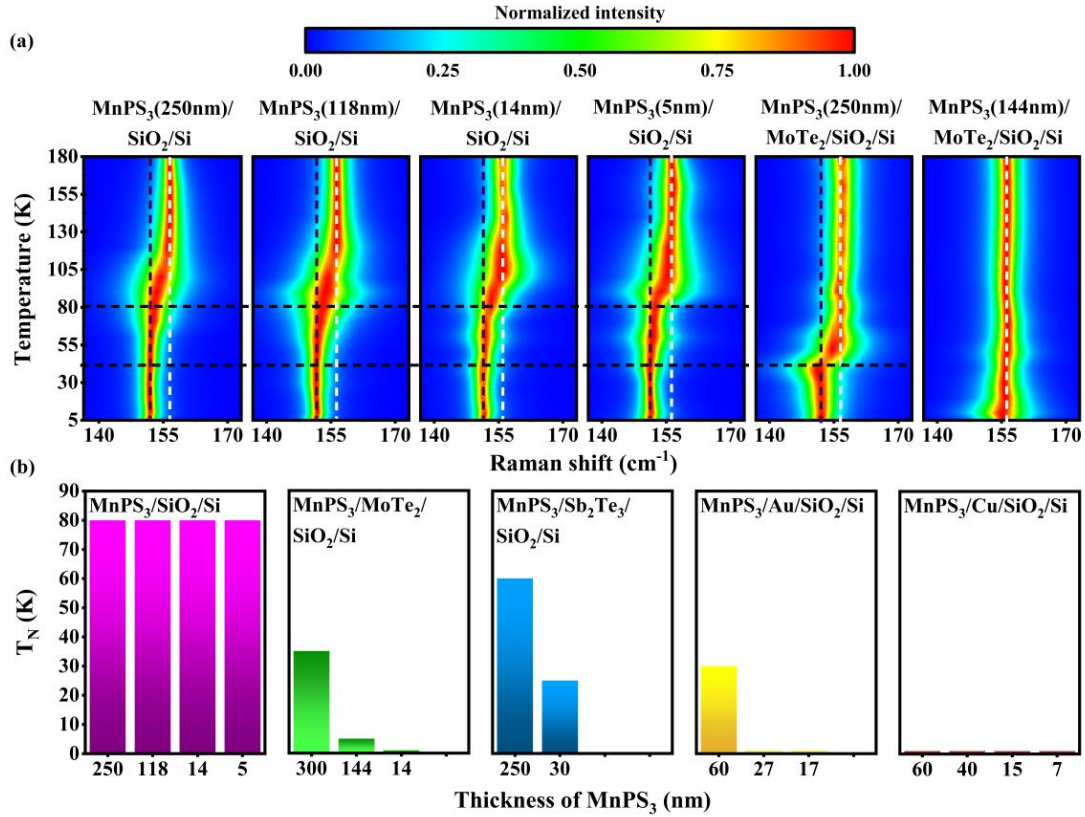

**Figure 2. (a)** The color maps show the intensity profile of the M$_2$ phonon across the magnetic phase transition for MnPS$_3$ flakes of various thickness on SiO$_2$/Si substrate and in heterostructures with T$_d$-MoTe$_2$. The dashed horizontal lines mark the corresponding Neel temperatures, **(b)** The obtained Neel temperatures of MnPS$_3$ in heterostructures of various thicknesses of MnPS$_3$ with varying underlayers/substrates (SiO$_2$/Si, MoTe$_2$, Sb$_2$Te$_3$, Au, and Cu).

i. **In MnPS$_3$:**

Fig. 2 shows the AFM ordering temperature for various flakes and heterostructures of MnPS$_3$, an isotropic Heisenberg magnet. As discussed previously, the M$_2$ phonon of MnPS$_3$ shifts from ~156 cm$^{-1}$ in the PM phase (represented by white vertical dashed lines) to ~151 cm$^{-1}$ in the AFM phase (represented by blue vertical dashed lines). From the contour maps in Fig. 2a, it can be clearly observed that the AFM ordering takes place at ~80 K for all the four flakes of MnPS$_3$ of different thicknesses (250, 118, 14, and 5 nm, respectively) supported on SiO$_2$/Si substrate (Note SN4 of Supplementary information). This is consistent with various recent experimental and theoretical reports[23,31]. The weak dependence of the T$_N$ on the flake thickness may be attributed to the weak interlayer coupling in MnPS$_3$[30]. We have investigated the heterostructures of MnPS$_3$ with MoTe$_2$, that transforms to the T$_d$ phase, a Weyl semimetallic phase, at low temperatures below ~250 K[33,34] (See Note SN5 in the Supplementary information). It is interesting to see that the proximity to the Weyl semimetallic T$_d$-MoTe$_2$ suppresses the T$_N$ of the MnPS$_3$ flakes considerably. Further, it is also observed that the suppression is different for different thicknesses of MnPS$_3$. The bulk exfoliated flake of MnPS$_3$ of thicknesses ~300 nm shows a suppression of the T$_N$ from ~80 K to ~35 K, while for the 144 nm thick flake, the T$_N$ shifts down to below 5 K, in the corresponding heterostructures with T$_d$-MoTe$_2$. For the MnPS$_3$ flake of thickness ~14 nm on T$_d$-MoTe$_2$, the Raman signatures are too weak to resolve due to its lower cross-section (not shown here). Nonetheless, the data on the ~14 nm thick flake suggest a complete suppression of the AFM phase. (See Note SN6 in the Supplementary information).

In order to understand the origin of the suppression, we have further performed experiments on MnPS$_3$ heterostructures with a range of different underlayers of topological and non-topological materials with a wide variety of SOC strengths and electronic density of states. The results obtained are summarized in Fig. 2b (also see Figure S10 of Supplementary information). While experiments performed on flakes of varying thickness directly on SiO$_2$/Si substrate showed no change from the reported bulk T$_N$ value ~80 K, strong suppression was observed for all the flakes supported on T$_d$-MoTe$_2$ (Type-II Weyl semimetal with strong SOC), Sb$_2$Te$_3$ (topological insulator with strong SOC), gold (metallic with strong atomic SOC), and copper (metallic with poor atomic SOC). Furthermore, the suppression was observed to be stronger for thinner flakes of MnPS$_3$.

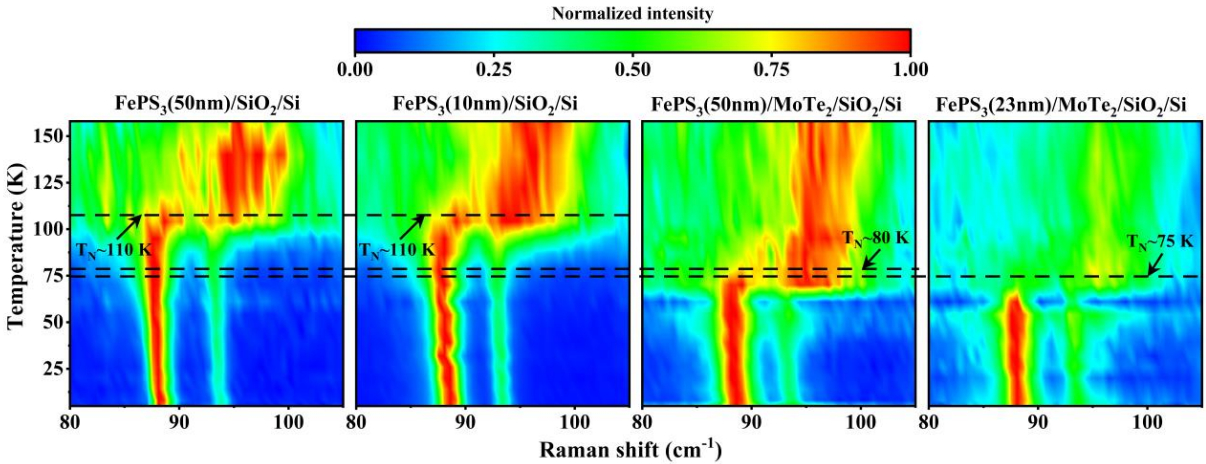

**Figure 3.** The color maps show the intensity profile of the F$_1$ phonon across the magnetic phase transition for FePS$_3$ flakes of various thickness on SiO$_2$/Si substrate and in heterostructures with T$_d$-MoTe$_2$. The dashed horizontal lines mark the corresponding Neel temperatures.

ii. **In FePS$_3$:**

Fig. 3 shows the contour maps corresponding to two different thicknesses of FePS$_3$ (Ising type AFM) on SiO$_2$/Si substrate and two heterostructures with T$_d$-MoTe$_2$ as underlayer. As discussed previously, the AFM ordering in FePS$_3$ is characterized by the emergence of four peaks (F$_{1a}$, F$_{1b}$, F$_{1c}$, and F$_{1d}$) from the F$_1$ mode and the drastic enhancement in the intensities of certain modes ($F_{1a}$, $F_{1b}$, $F_{1c}$, $F_{1d}$, and $F_2$) below the $T_N$. Though all these signatures were witnessed in our investigated flakes at low temperatures (See Fig. S11 in the Supplementary information for more details), we have shown only the most prominent ones ($F_{1a}$ and $F_{1b}$) in the contour map in Fig. 3. It can be clearly observed that the $T_N$ (temperature were the $F_{1a}$ and $F_{1b}$ features disappear and the asymmetric $F_1$ mode appears instead) for $FePS_3$ flakes of thickness ~50 nm and ~10 nm on $SiO_2$/Si substrate are measured to be ~110 K, again pointing out its insensitivity to the flake thickness. The heterostructures of $FePS_3$ and $T_d$-$MoTe_2$ with two different thicknesses (~50 nm and ~23 nm) of the $FePS_3$ layers exhibited a suppression of the $T_N$ to ~80 K for the 50 nm thick flake of $FePS_3$ and to ~75 K for the 23 nm flake on $T_d$-$MoTe_2$. This is consistent with the observation of suppression of $T_N$ in $MnPS_3$.

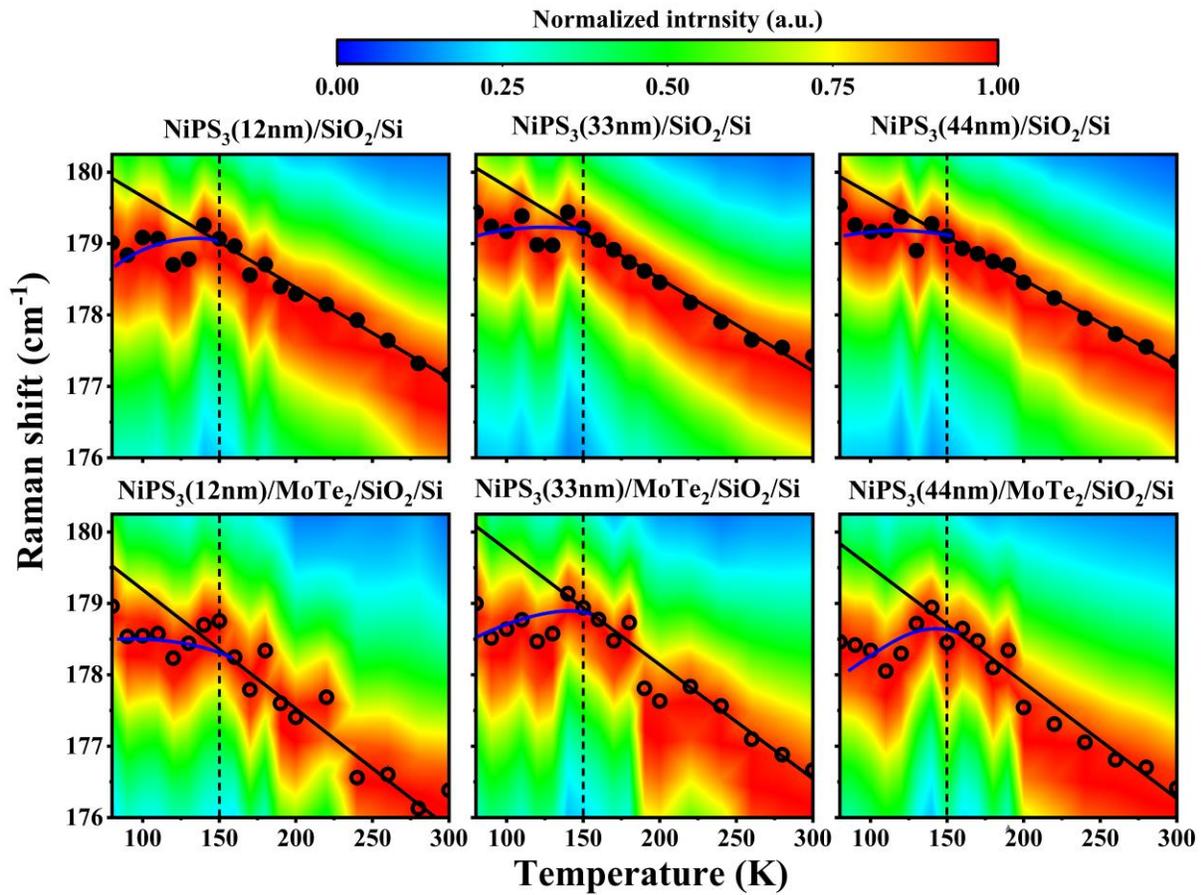

**Figure 4.** The color maps show the intensity profile of the $N_2$ phonon across the magnetic phase transition for $NiPS_3$ flakes of various thickness on $SiO_2$/Si substrate and in heterostructures with $T_d$-$MoTe_2$. The $N_2$ phonon clearly shows deviations (represented by solid blue line) from anharmonic trend (represented by solid black lines). The onsets of the deviations are measured as the Neel temperatures for the corresponding flakes of $NiPS_3$ and are marked by the dashed vertical lines.

iii.     **In $NiPS_3$:**

Figure 4 shows the temperature dependence of the N$_2$ mode of NiPS$_3$ (an XY-type AFM), for various flakes and corresponding heterostructures with T$_d$-MoTe$_2$. We observe that irrespective of the flake thickness, the mode exhibits a departure from the anharmonic behavior (represented by the black solid line) below ~150 K. This behavior (deviation from the anharmonic trend), as discussed previously, is accounted for the onset of spin-phonon coupling associated with the AFM spin ordering, providing a measure of the T$_N$ for NiPS$_3$ flakes. (Other signatures of the AFM phase transition are shown in Figures S12 and S13 of Supplementary information). Therefore, flakes of thicknesses 12, 33, and 44 nm supported on SiO$_2$/Si substrate, exhibit T$_N$ of ~150 K, implying its invariance with the thickness. It is interesting to observe that unlike MnPS$_3$ and FePS$_3$, the introduction of T$_d$-MoTe$_2$ underlayers show no considerable changes in the magnetic ordering temperature (T$_N$) with respect to the flakes directly on the SiO$_2$/Si substrate. The observations on NiPS$_3$ are in sharp contrast to those in the case of MnPS$_3$/FePS$_3$ discussed above.

**Discussion**

Our primary observations are: (i) the strong suppression of the AFM phase of MnPS$_3$ and FePS$_3$ in their heterostructures with T$_d$-MoTe$_2$ which are otherwise robust to their flake thickness, while (ii) the AFM phase in NiPS$_3$ remains robust even in its heterostructures with T$_d$-MoTe$_2$. In this section, we discuss these findings and their possible origin. Chittari *et al.*[35] using *ab initio* calculations have predicted a transition from semiconducting AFM to metallic FM phase in MPS$_3$ compounds by means of charge doping and application of strain. These possibilities may be ruled out in our experiments based on various evidences as explained in detail in Supplementary information (Note SN8 and Figures S16 and S17). Instead, we argue that DM interaction arising from the substrate is the principal driving mechanism.

The magnetic interactions between neighboring spins of the MPS$_3$ compounds may be described by the following Hamiltonian[13]:

$$H = -2\sum\{J_\perp(S_{ix}S_{jx} + S_{iy}S_{jy}) + J_\parallel(S_{ix}S_{jx} + S_{iy}S_{jy}) + AS_{iz}^2\} + H_{DM} \qquad \ldots (1)$$

$$H_{DM} = \boldsymbol{D_{ij}} \cdot (\boldsymbol{S_i} \times \boldsymbol{S_j}) \qquad \ldots (2)$$

Here $J_\perp$ and $J_\parallel$ represent the perpendicular and parallel exchange interactions between the spins $\boldsymbol{S_i}$ and $\boldsymbol{S_j}$, while $A$ represents the single-ion anisotropy induced by the axial distortion. The last term in equation (1) captures the Dzyaloshinskii Moriya (DM) interaction and is expanded in equation (2). While the first two terms support a collinear arrangement of neighboring spins, the $H_{DM}$ prefers a perpendicular arrangement[11,36,37]. However, a non-zero DM interaction requires SOC as well as absence of inversion symmetry.

The bulk MPS$_3$ (M=Mn, Fe, and Ni) crystals, on account of their monoclinic structures, preserve the inversion symmetry and, therefore, DM interactions are absent. However, the exfoliated thin films of MPS$_3$ break the inversion symmetry at their interfaces and are ideal platforms to introduce DM interactions[11]. In particular, the breaking of inversion symmetry at the interface of a heterojunction can give rise to momentum-dependent spin split dispersions as a consequence of Dresselhaus or Rashba type SOC[11,38], which, in turn lead to significant DM interactions at the interfaces of magnetic heterojunctions[39,40] particularly in presence of low energy charge carriers, *i.e.*, in metallic and semimetallic substrate. Indeed, we see a more pronounced suppression of magnetic transition temperature in such cases (*e.g.*, $T_d$-$MoTe_2$, $Sb_2Te_3$, Au, and Cu) compared to the insulating substrate such as $SiO_2$/Si. Further, we also observe that despite having similar thicknesses (~60 nm), the $MnPS_3$ flake on Cu shows stronger suppression of AFM ordering compared to a similarly thick flake on Au. This, we attribute to the higher density of states associated with Cu as compared to Au[41]. Given the substantial suppression on Cu substrate (with negligible atomic SOC) as well as Au (with substantial atomic SOC), we think the atomic spin-orbit coupling for the electrons in the substrate plays a secondary role in determining the strength of the resultant DM interactions. It is important to note that the DM interactions would be dominant for the spins in the vicinity of the interface with the non-magnetic layer and not necessarily in the entire bulk of the flake. Notably, we can rule out the contribution of RKKY effect, mediated through free carriers, on the DM interactions due to the wide-gap semiconducting nature of $MPS_3$[35]. Interestingly, few recent studies have reported the magnetic properties and ordering temperatures in AFMs to be strongly affected by modifications of the surface magnetism[42–44]. The DM interactions in our experiments similarly affect the magnetic ordering at the interfacial surface of the magnetic layer, which in turn is likely to affect the magnetic ground state of the entire bulk flake. An exact calculation of the surface magnetism effects on the magnetic properties is crucial to shed further light on the mechanism which is beyond the scope of this work.

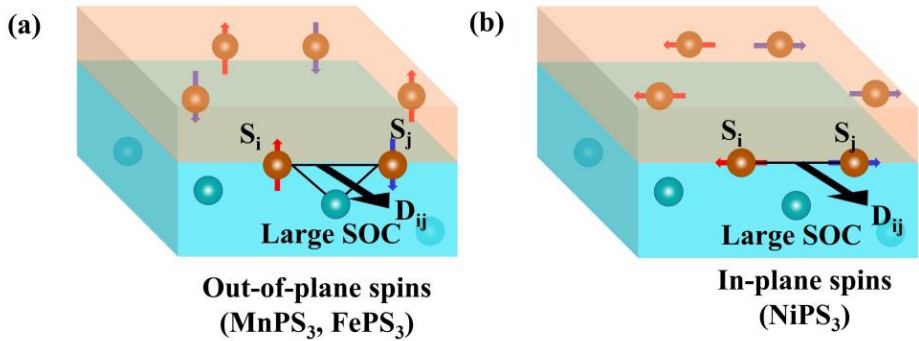

**Figure 5.** Schematic representation of the Dzyaloshinskii Moriya interactions in **(a)** magnetic thin films with spins oriented out-of-plane (*e.g.*, $MnPS_3$ $FePS_3$) and **(b)** magnetic thin films with spins oriented in-plane (*e.g.*, $NiPS_3$).

Indeed, recent ARPES measurements revealed that Au and Cu possess Rashba SOC[45,46]. Therefore, in all our measurements on the thin flakes of $MPS_3$ (M=Mn, Fe, and Ni) supported on underlayers/substrates of $T_d$-$MoTe_2$, $Sb_2Te_3$, Au, and Cu, due to the broken inversion symmetry and presence of SOC, we must observe a contribution of DM interactions. This is very significant as the DM interaction is supposed to be dominant at the vicinity of the interface (where the inversion symmetry is broken), thereby, introducing a plausible dependence of the magnetic interactions on the flake thickness. The competition between Heisenberg and DM interaction terms induces a canting of the perfectly anti-parallel spins in an AFM (Fig. 5a), resulting in weakening of the AFM ordering, thereby, explaining the suppression of the AFM phase in the investigated MPS$_3$ thin films. Additionally, strong DM interactions can also result in topologically protected spin textures like skyrmions and chiral domain walls[6–10] with novel spin-orbitronic and storage applications[11,12]. Our results also bring out certain traits of the interfacial DM interactions observed in the 2D magnetic MPS$_3$ thin films, as explained next.

**Effect of spin orientation associated with the magnetic layer:** The DM interaction energy between two spins $S_i$ and $S_j$, as described by equation (2), clearly invokes a dependence on the spin orientations. For the heterostructures, the DM vector $D_{ij}\,(=\,-D_{ji}) = n \times r_{ij}$ lies in the plane of the interface between the magnetic and the non-magnetic layer while $n$ and $r_{ij}$ being the normal to the interface and the separation between the spins, $S_i$ and $S_j$, respectively[11]. It is evident from the above expression that the DM interactions would be drastically different for the out-of-plane and the in-plane spins, as illustrated schematically in Fig. 5. While for the out-of-plane spins, the DM interactions result in the canting of the spins towards each-other along a direction perpendicular to their separation ($r_{ij}$) and the normal ($n$) to the surface, the DM interactions would vanish for the in-plane spins (which are coplanar with the $D_{ij}$ vector). This explains why the introduction of T$_d$-MoTe$_2$ suppresses the AFM ordering temperature for MnPS$_3$ and FePS$_3$ (spins oriented out-of-plane), while NiPS$_3$ (spins oriented in-plane) remains unaffected. Among MnPS$_3$ and FePS$_3$, the suppression of the AFM phase is stronger in the case of MnPS$_3$ plausibly because MnPS$_3$ exhibits an isotropic Heisenberg type interaction ($J_\parallel = J_\perp$), while FePS$_3$ shows an anisotropic behavior ($J_\parallel \neq J_\perp$)[13]. Therefore, randomization of the spins in MnPS$_3$ would cost lesser energy as compared to FePS$_3$, which has an out-of-plane anisotropic axis. As opposed to the results obtained for MnPS$_3$ and FePS$_3$, we observed that NiPS$_3$ flakes (which have their spins oriented in-plane) showed no appreciable changes in the T$_N$ in the heterostructures with T$_d$-MoTe$_2$.

**Conclusion**

In summary, we have prepared heterostructures of MPS$_3$ (M=Mn, Fe, and Ni) compounds with T$_d$-MoTe$_2$, Sb$_2$Te$_3$, Au, as well as Cu, and observed strong suppression of magnetic ordering, which have been attributed to the presence of strong DM interactions. The DM interactions have been demonstrated to show strong sensitivity towards the spin orientation in the magnetic material, showing strongest effects for the out-of-plane spins and no effect on the in-plane spins. The experiments performed on MPS$_3$ supported on a variety of substrates with varying DOS also suggest a strong dependence of the strength of DM interactions on the electronic density of states associated with the underlayers/substrates. We believe that our extensive study will be vital to engineer new spin textures for potential spintronic applications.

**Methods**

1. *Synthesis:* The crystals of Mn/NiPS$_3$ and Sb$_2$Te$_3$ were procured commercially (HQ Graphene).
a. **MoTe₂:** Single crystals of 1T′-MoTe$_2$ were grown by chemical vapor transport method using iodine as the transport agent. In the first step, stoichiometric mixtures of Mo (99.9% pure) and

Te (99.99% pure) powders were ground together, pelletized, and sealed in an evacuated quartz tube. The sealed ampoule was first heated at 1100 ºC for 24 h, followed by ice water quenching to avoid formation of the 2$H$ phase and the same heat treatment was repeated. The obtained phase-pure polycrystalline sample was used in the crystal growth process. Crystallization was carried out from hot zone (1100 ºC) to cold zone (900 ºC) in a two-zone furnace where the polycrystalline sample was kept in the hot zone. Finally, the sealed quartz tube was quenched in air to avoid the formation of the hexagonal phase.

b. **FePS$_3$:** Single crystals of FePS$_3$ were grown using the physical vapor transport technique. Thoroughly mixed high-purity elemental powders taken in a stoichiometric ratio were loaded in a pre-heated quartz ampoule and sealed under a 10$^{-5}$ mbar vacuum. Subsequently, we placed the ampoule in an oven with independently controlled heating zones. Simultaneously, the source and sink zone temperatures were raised slowly to 750 ºC, respectively 730 ºC, and maintained under this temperature gradient for ten days, followed by slow cooling to room temperature. Multiply shiny-platelets single crystals with lateral dimensions exceeding, in some cases, 10 mm were observed in the sink zone.

2. *Preparation of Cu and Au substrates:*
a. **Thermal evaporation:** Copper was deposited on SiO$_2$/Si substrate by thermal evaporation technique.
b. **Sputtering:** Gold was deposited on SiO$_2$/Si substrate by sputtering technique.
3. *Preparation of heterostructures:* The heterostructures were prepared by a dry transfer technique using a PDMS film. The flakes were micromechanically exfoliated on to a transparent PDMS film, then by positioning the layers suitably under a microscope, the layers were transferred on top of the pre-transferred underlayer (also prepared by micromechanical exfoliation) on SiO$_2$/Si substrate in order to prepare the required heterostructures.
4. *Determination of thickness of the flakes:*
   **Atomic force microscopy:** The thickness of the various flakes and heterostructure investigated were confirmed through AFM measurements performed in a Bruker Dimension ICON PT atomic force microscope in non-contact mode.
5. *Raman measurements:*
   The Raman measurements were performed using a Horiba JY LabRam HR Evolution Raman spectrometer in back scattering geometry. A 532 nm (Nd:YAG-diode) laser was used as the excitation source, while the collection was done through a grating of 1800 grooves/mm. Temperature-dependent experiments were performed using two different cryostats. All the measurements were performed using low laser power (<1 mW) to avoid local heating effects.
a. For high temperature experiments, in the range of 80 – 300 K, a liquid N$_2$ based cryostat from LINKAM was used. The spectra were collected using a 50× objective (N.A. = 0.5).
b. For the low-temperature experiments between 5 – 200 K, a closed-cycle He cryostat (AttoDRY 1000) attached with a 100× objective lens (N.A.=0.82) was used.

**Acknowledgements**

S. Saha acknowledges funding from the Science and Engineering Research Board (SERB), India (Grants no. ECR/2016/001376 and no. CRG/2019/002668) and Ministry of Education, India (Grant no. STARS/APR2019/PS/662/FS). D. Negi acknowledges CSIR for fellowship (09/1020/(0139)/2018-EMR-I). Authors acknowledge Physics PG laboratory and Central Instrumentation Facility (IISER Bhopal) for thermal evaporation and sputtering facilities. L. Harnagea acknowledges financial support from DST-India (DST/WOS-A/PM-83/2021 (G)) and IISER Pune for providing the facilities for crystal growth and physical characterization. R. P. Singh acknowledges fundings from DST/SERB (Grant No. YSS/2015/001799). A. Soni acknowledges IIT Mandi for research facilities. S. Bhattacharjee acknowledges the adjunct fellow program at SNBNCBS, Kolkata, and funding from Max Planck partner group grant at ICTS, SERB-DST (India) for Swarna Jayanti grant SB/SJF/2021- 22/12-G and the Department of Atomic Energy, Government of India, under Project No. RTI4001.

**Author contribution statement**

S. Paul and S. Saha conceived the idea. N. Pistawala and L. Harnagea were involved in the synthesis of the $FePS_3$ single crystals, while M. Mandal, S. Marik, and R. P. Singh. were involved in the synthesis of the $MoTe_2$ crystal. The exfoliated thin films of various $MPS_3$ compounds and their heterostructures were prepared by S. Paul, D. Negi, and S. Talukdar. Raman measurements were performed by S. Paul, D. Negi, S. Talukdar, S. Karak, and B. Poojitha. Atomic force microscopy measurements were performed by A. Thomas and A. Soni. All data analysis and data interpretation were done by S. Paul. Magnetic measurements were performed by S. Paul, D. Negi, and S. Badola. The manuscript was written by S. Paul, S. Saha, and S. Bhattacharjee in consultation with all the authors.

**Competing financial interests:** The authors declare no competing financial interests.

# Supporting information

## SN1. Magnetization measurements

The magnetic properties of the bulk single crystals of $MnPS_3$, $FePS_3$, and $NiPS_3$ were measured in a Quantum Design Magnetic Properties Measurement System in vibrating sample magnetometer (VSM) mode. Figure S1 shows the magnetic susceptibility as a function of temperature for the $MnPS_3$, $FePS_3$, and $NiPS_3$ crystals. The corresponding Neel temperatures ($T_N$) have been measured from the broad maxima in the susceptibility ($\chi$) plots and also from the peaks observed in the $\frac{d\chi}{dT}$ (shown in insets). The measured $T_N$ values for $MnPS_3$, $FePS_3$, and $NiPS_3$ are ~80 K, ~115 K, and ~150 K, respectively. The measured values are consistent with prior reports[13].

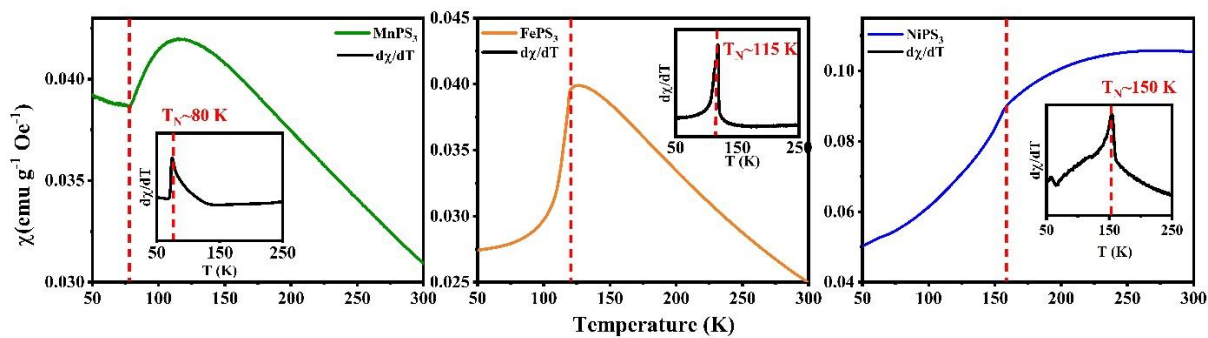

**Figure S1.** The temperature variation of magnetic susceptibilities ($\chi$) and the respective temperature derivatives $\left(\frac{d\chi}{dT}\right)$ (shown in insets) for $MnPS_3$, $FePS_3$, and $NiPS_3$ single crystals. The Neel temperatures ($T_N$) for the crystals are labelled with red vertical dashed lines.

## SN2. Heterostructures of MPS$_3$ investigated

We have investigated various flakes of MPS$_3$ (M=Mn, Fe, and Ni) of varying thicknesses by preparing a series of heterostructures, which can be broadly classified into the following categories:

Type 1: MnPS$_3$ flakes of various thicknesses on SiO$_2$/Si substrate
Type 2: FePS$_3$ flakes of various thicknesses on SiO$_2$/Si substrate
Type 3: NiPS$_3$ flakes of various thicknesses on SiO$_2$/Si substrate
Type 4: Heterostructures of MnPS$_3$ and MoTe$_2$ on SiO$_2$/Si substrate
Type 5: Heterostructures of FePS$_3$ and MoTe$_2$ on SiO$_2$/Si substrate
Type 6: Heterostructures of NiPS$_3$ and MoTe$_2$ on SiO$_2$/Si substrate
Type 7: Heterostructures of MnPS$_3$ and Sb$_2$Te$_3$ on SiO$_2$/Si substrate
Type 8: MnPS$_3$ flakes of various thicknesses on Au-coated SiO$_2$/Si substrate
Type 9: MnPS$_3$ flakes of various thicknesses on Cu-coated SiO$_2$/Si substrate

Figure S2 shows the schematics of each of these categories. We have prepared several heterostructures of each type (Type 1 - 9) by varying the thickness of the top magnetic flake (MnPS$_3$/ FePS$_3$/ NiPS$_3$).

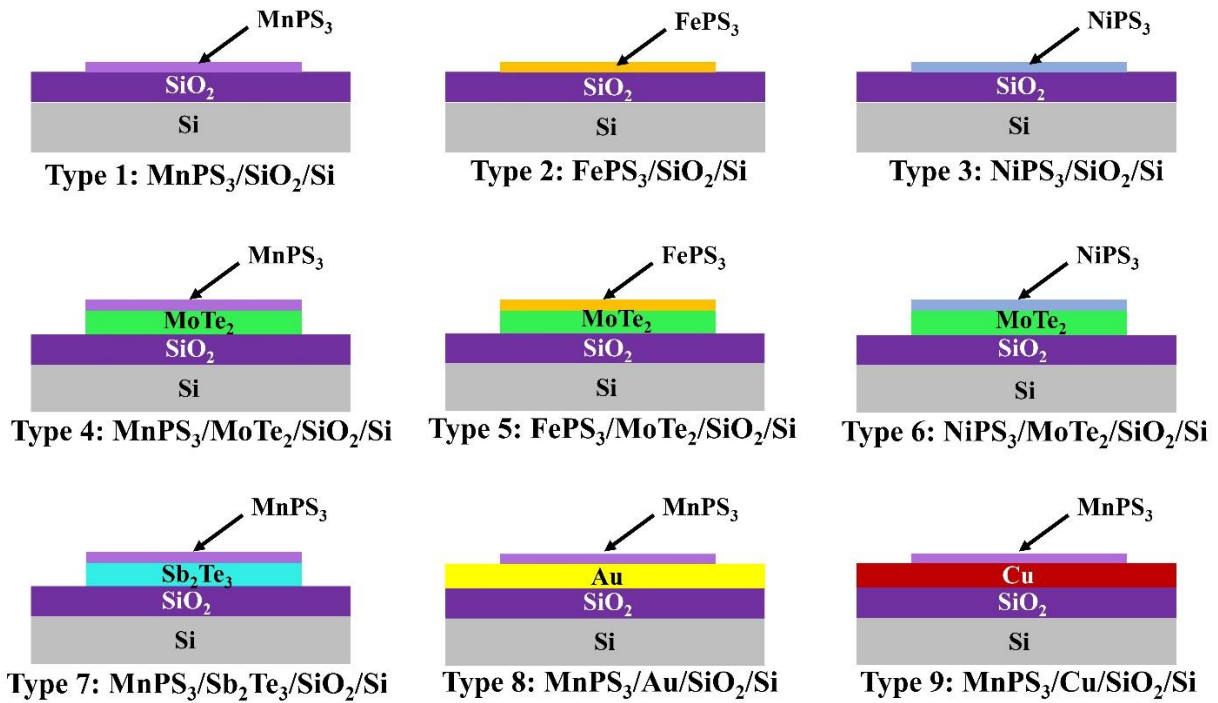

**Figure S2.** The schematic representation of the nine different types of heterostructures investigated in the present study.

## SN3. Determination of flake thicknesses

The magnetic properties of certain 2D materials have been reported to strongly depend on the flake thickness of the material. In our measurements, we have observed that the magnetic properties (particularly the Neel temperature) of the $MPS_3$ compounds to show robustness to the flake thickness. However, under the influence of proximity to a suitable underlayer/substrate in the heterostructures, we observe an induction of a systematic dependence of the Neel temperature on the flake thickness. Therefore, it is of vital importance to determine the thickness of the flakes of the magnetic layer in the various samples that were studied. In this regard, we have performed atomic force microscopy on the various investigated samples. Figures S3-S9 show the optical images (obtained using a 50× objective lens) and the corresponding atomic force microscopy profiles for the samples (with different thicknesses of $MPS_3$ layers) corresponding to the various types of heterostructures (Type 1 - Type 9), enlisted in Supplementary note SN2.

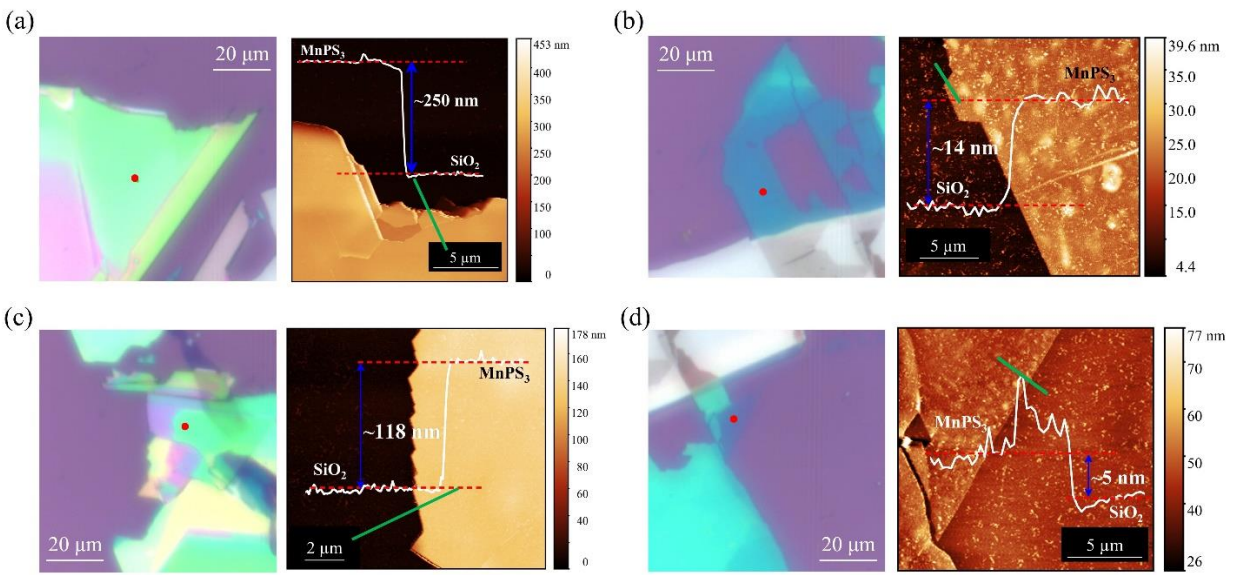

**Figure S3. The optical images and the corresponding AFM topographies of the samples of Type 1 ($MnPS_3/SiO_2/Si$). The red dots in the optical images mark the spots that were investigated. The thickness of the $MnPS_3$ flakes (measured from the step height along the green lines on the AFM profiles) were obtained as (a) ~250 nm, (b) ~14 nm, (c) ~118 nm, and (d) 5 nm, respectively.**

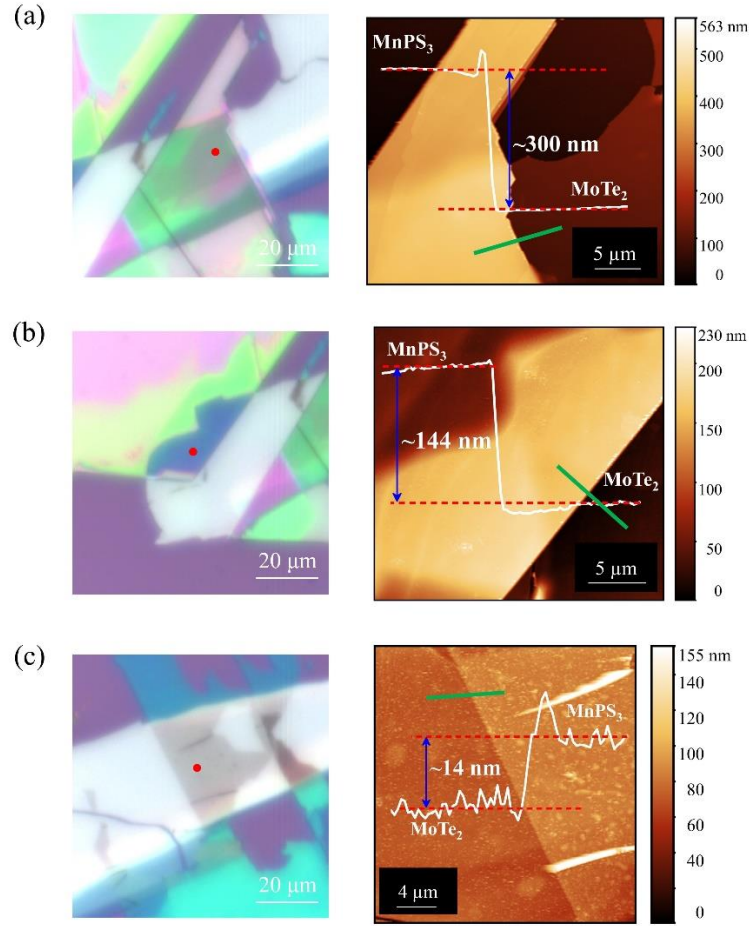

**Figure S4.** The optical images and the corresponding AFM topographies of the samples of Type 4 (MnPS$_3$/MoTe$_2$/SiO$_2$/Si). The red dots in the optical images mark the spots that were investigated. The thickness of the MnPS$_3$ flakes (measured from the step height along the green lines on the AFM profiles) were obtained as (a) ~300 nm, (b) ~144 nm, and (c) ~14 nm, respectively.

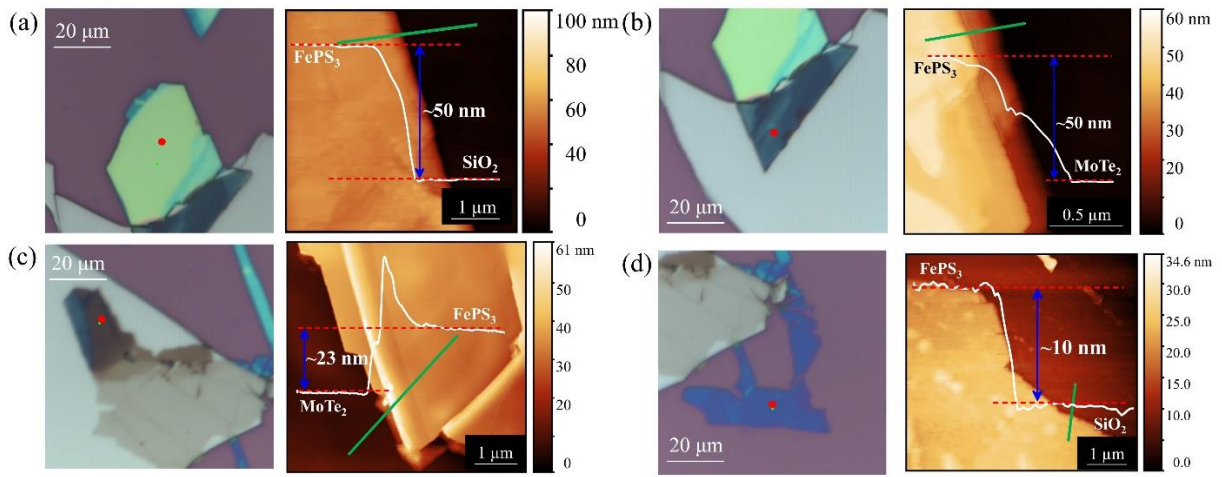

**Figure S5.** The optical images and the corresponding AFM topographies of the samples of Type 2 (FePS$_3$/SiO$_2$/Si) and Type 5 (FePS$_3$/MoTe$_2$/SiO$_2$/Si). The red dots in the optical images mark the spots that were investigated. The thickness of the FePS$_3$ flakes (measured from the step height along the green lines on the AFM profiles) were obtained as (a) ~50 nm, (b) ~50 nm, (c) ~23 nm, and (d) ~10 nm, respectively.

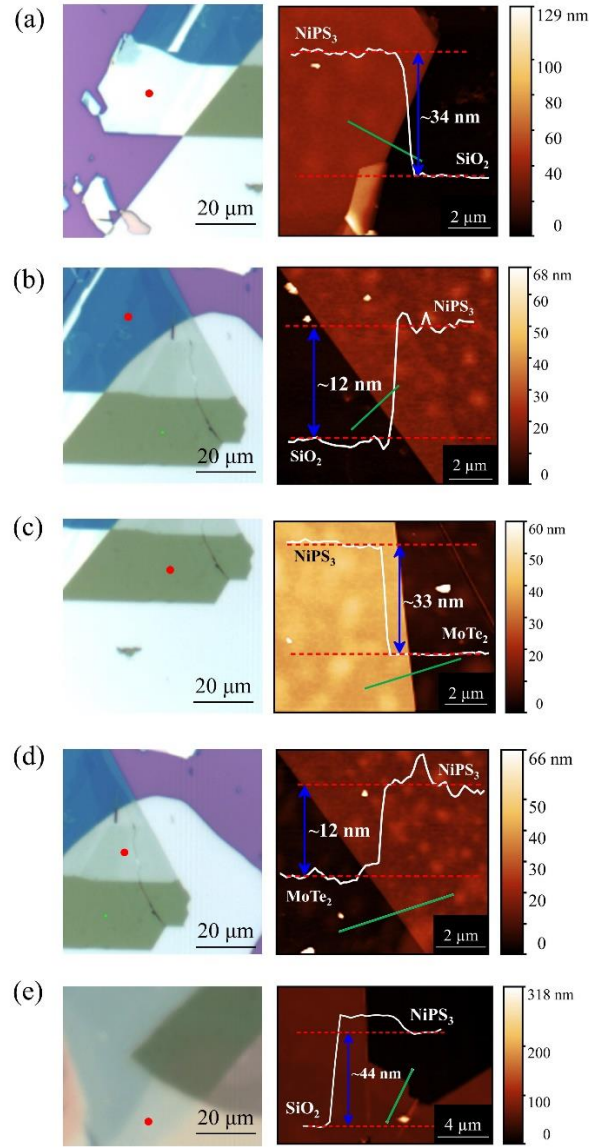

**Figure S6.** The optical images and the corresponding AFM topographies of the samples of Type 3 (NiPS$_3$/SiO$_2$/Si) and Type 6 (NiPS$_3$/MoTe$_2$/SiO$_2$/Si). The red dots in the optical images mark the spots that were investigated. The thickness of the NiPS$_3$ flakes (measured from the step height along the green lines on the AFM profiles) were obtained as (a) ~34 nm, (b) ~12 nm, (c) ~33 nm, (d) ~12 nm, and (e) ~44 nm, respectively.

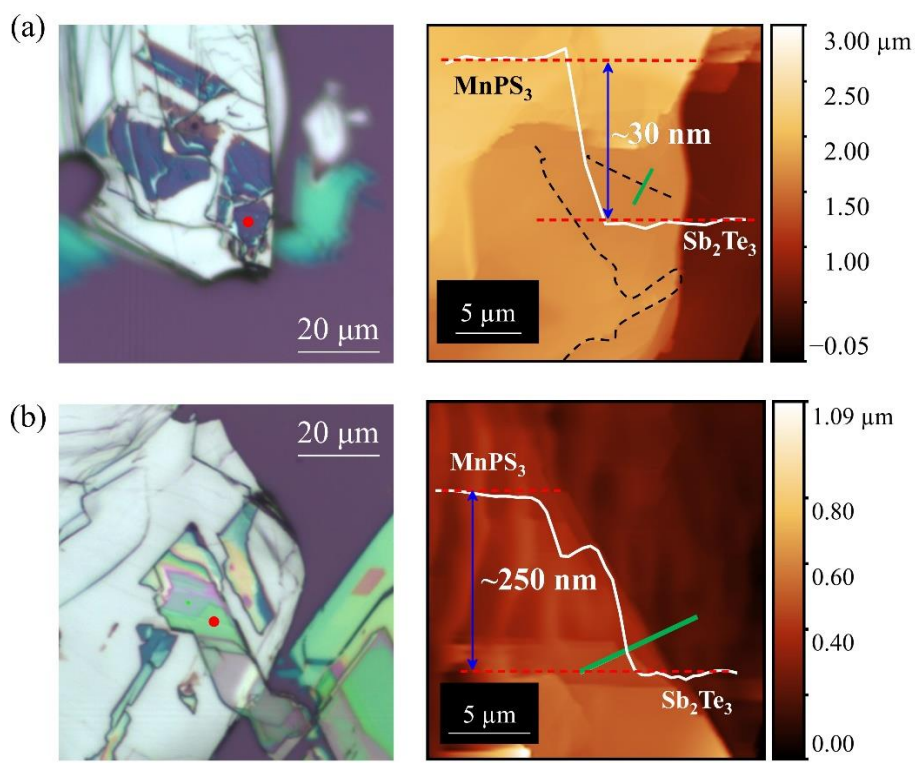

**Figure S7.** The optical images and the corresponding AFM topographies of the samples of Type 7 (MnPS$_3$/Sb$_2$Te$_3$/SiO$_2$/Si). The red dots in the optical images mark the spots that were investigated. The thickness of the MnPS$_3$ flakes (measured from the step height along the green lines on the AFM profiles) were obtained as (a) ~30 nm and (b) ~250 nm, respectively.

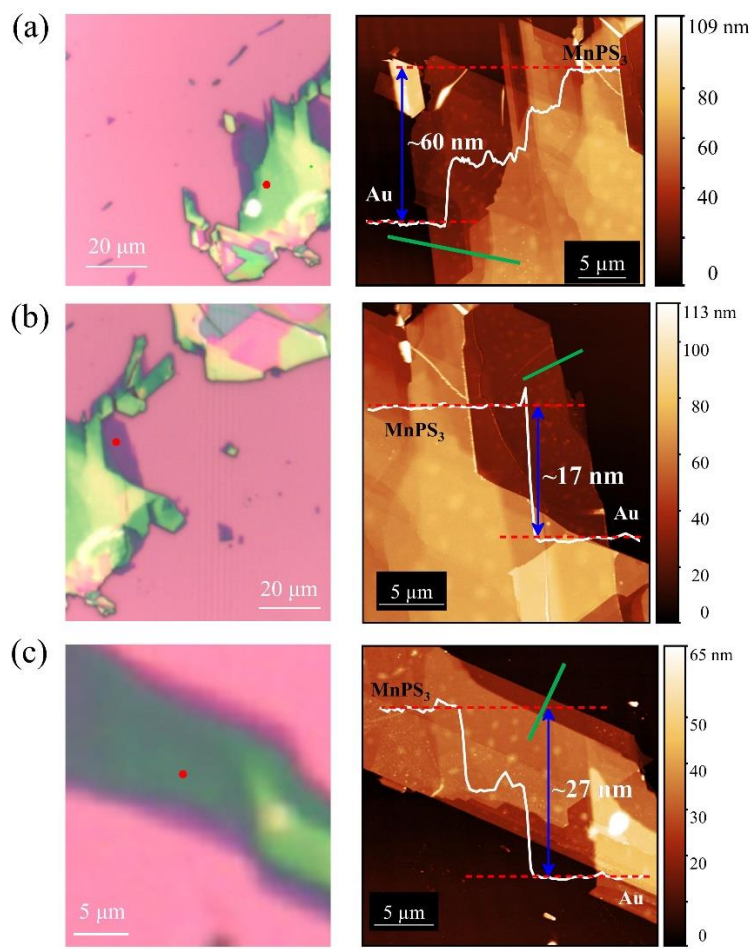

**Figure S8.** The optical images and the corresponding AFM topographies of the samples of Type 8 (MnPS$_3$/Au/SiO$_2$/Si). The red dots in the optical images mark the spots that were investigated. The thickness of the MnPS$_3$ flakes (measured from the step height along the green lines on the AFM profiles) were obtained as (a) ~60 nm, (b) ~17 nm, and (c) ~27 nm, respectively.

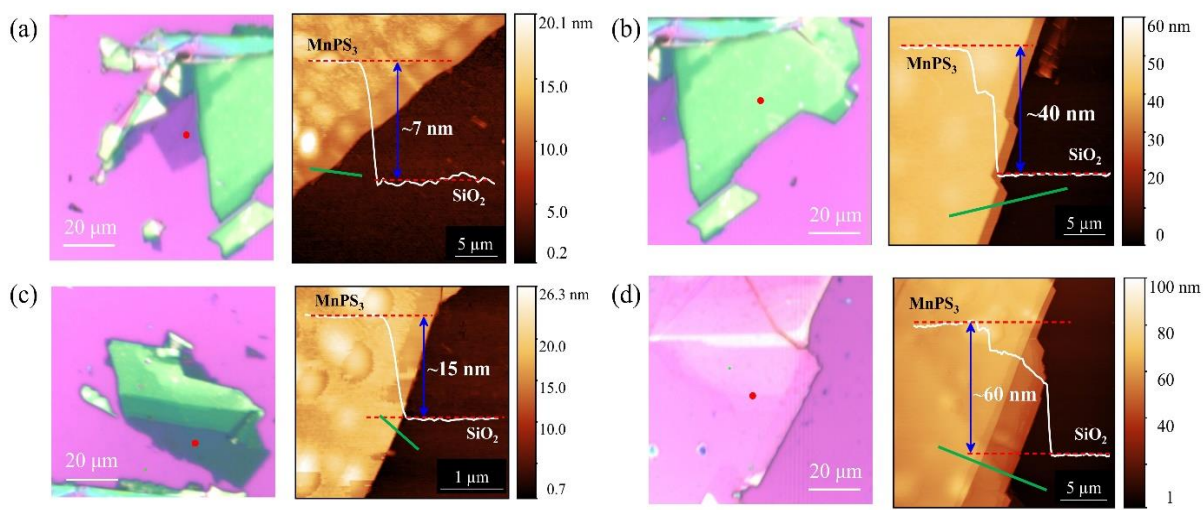

**Figure S9.** The optical images and the corresponding AFM topographies of the samples of Type 9 (MnPS$_3$/Cu/SiO$_2$/Si). The red dots in the optical images mark the spots that were investigated. The thickness of the MnPS$_3$ flakes (measured from the step height along the green lines on the AFM profiles) were obtained as (a) ~7 nm, (b) ~40 nm, (c) ~15 nm, and (c) ~60 nm, respectively.

## SN4. Detection of Neel temperature

The antiferromagnetic (AFM) ordering in the various investigated samples have been probed using Raman spectroscopy. As explained in the main text, there are various signatures in the Raman spectra for all the MPS$_3$ (M= Mn, Fe, and Ni) compounds which may be used for confirmation of the appearance of the AFM phase in the respective samples and hence to determine the corresponding Neel temperatures. The strongest signatures associated with the AFM phase are generally observed in the temperature dependence of the low frequency phonons which are associated with the vibration of the magnetic ions (Mn$^{2+}$, Fe$^{2+}$. and Ni$^{2+}$). While the most prominent signatures have been used in the main text for the determination the Neel temperatures for the various samples, we will show some related results here, which further confirm the accurate determination of the Neel temperatures.

### (a) MnPS$_3$

As discussed in the main text, the M$_2$ phonon in the Raman spectra of MnPS$_3$ shows strong signature of the magnetic transition from the paramagnetic (PM) to the AFM phase. The M$_2$ phonon undergoes an anomalous redshift as the temperature is lowered below the Neel temperature, resulting in a step-like behavior in its temperature-dependence. Therefore, the evolution of the M$_2$ phonon across the magnetic ordering temperature is used to accurately measure the Neel temperature of MnPS$_3$. Figure S10 shows the temperature evolution of the frequency of the M$_2$ phonon for all the MnPS$_3$ flakes in the various heterostructures studied, and the corresponding Neel temperatures are also labeled, wherever applicable.

### (b) FePS$_3$

In case of FePS$_3$, the phonon F$_1$ splits into four components, F$_{1a}$, F$_{1b}$, F$_{1c}$, and F$_{1d}$ as the sample is cooled below the Neel temperature. Further, the mode F$_2$ narrows down with a simultaneous increase in the peak height. Therefore, these behaviors exhibited by the modes F$_1$ and F$_2$ is used as a Raman probe for the detection of the magnetic transition (Figure S11a). The signature is particularly strong for the mode F$_{1a}$ which exists in the AFM phase and vanishes in the PM phase. Therefore, the intensity (or height) of the F$_{1a}$ phonon has been used in our experiments to determine the Neel temperature of the FePS$_3$ flakes. Figure S11b shows the temperature-dependent peak height of the F$_{1a}$ phonon across the phase transition and the corresponding Neel temperatures have been measured for the various investigated heterostructures of FePS$_3$.

### (c) NiPS$_3$

For NiPS$_3$, there are several signatures associated with the transition from the PM to the AFM phase. The main text shows the signature of spin-phonon coupling exhibited by the N$_2$ phonon, the onset of which marks the Neel temperature. However, some other typical signatures are also observed. Firstly, a broad 2M feature appears in the magnetically ordered phase, the height of which drastically goes to nearly zero above the Neel temperature. Further, the N$_2$ phonon constitutes of two components which are degenerate in the PM phase, but below the T$_N$, the degeneracy is lifted. Figure S12 and S13 show these signatures, respectively, to further confirm an accurate estimation of the Neel temperature of the various flakes of NiPS$_3$.

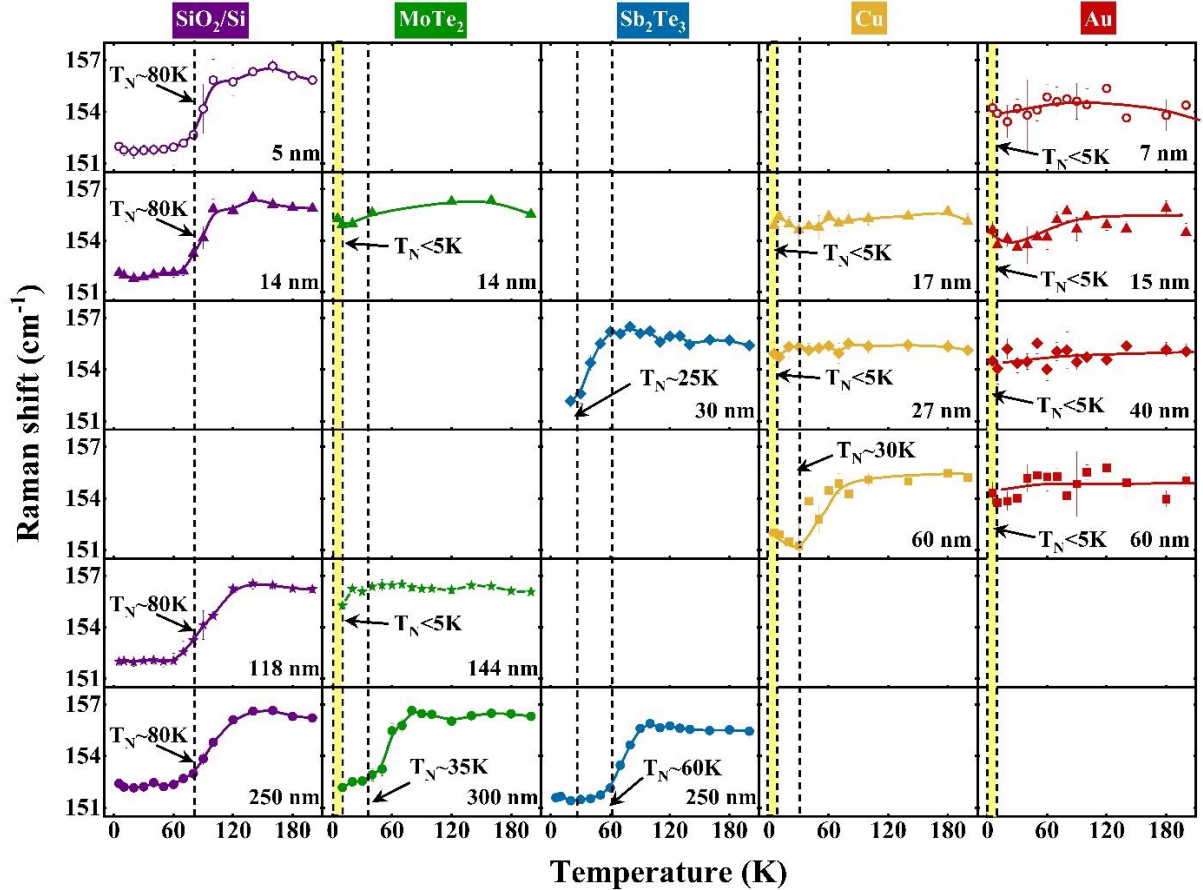

Figure S10. The temperature evolution of the mode frequency of the $M_2$ phonon for the various $MnPS_3$ flakes on different underlayers/substrates. The corresponding Neel temperatures are labeled. The samples which do not show any signature of transition to the AFM phase down to 5 K (the lowest measured temperature in our experiments) have been labelled to have their $T_N$ < 5K (represented by yellow boxes). Each row (and symbol) represents data of $MnPS_3$ flakes of similar thickness but supported on different underlayer/substrates. Each column represents data corresponding to flakes of $MnPS_3$ of different thicknesses on the same underlayer/substrate. We clearly observe a suppression of the Neel temperatures in all the heterostructures with respect to the flakes directly on $SiO_2/Si$ substrate.

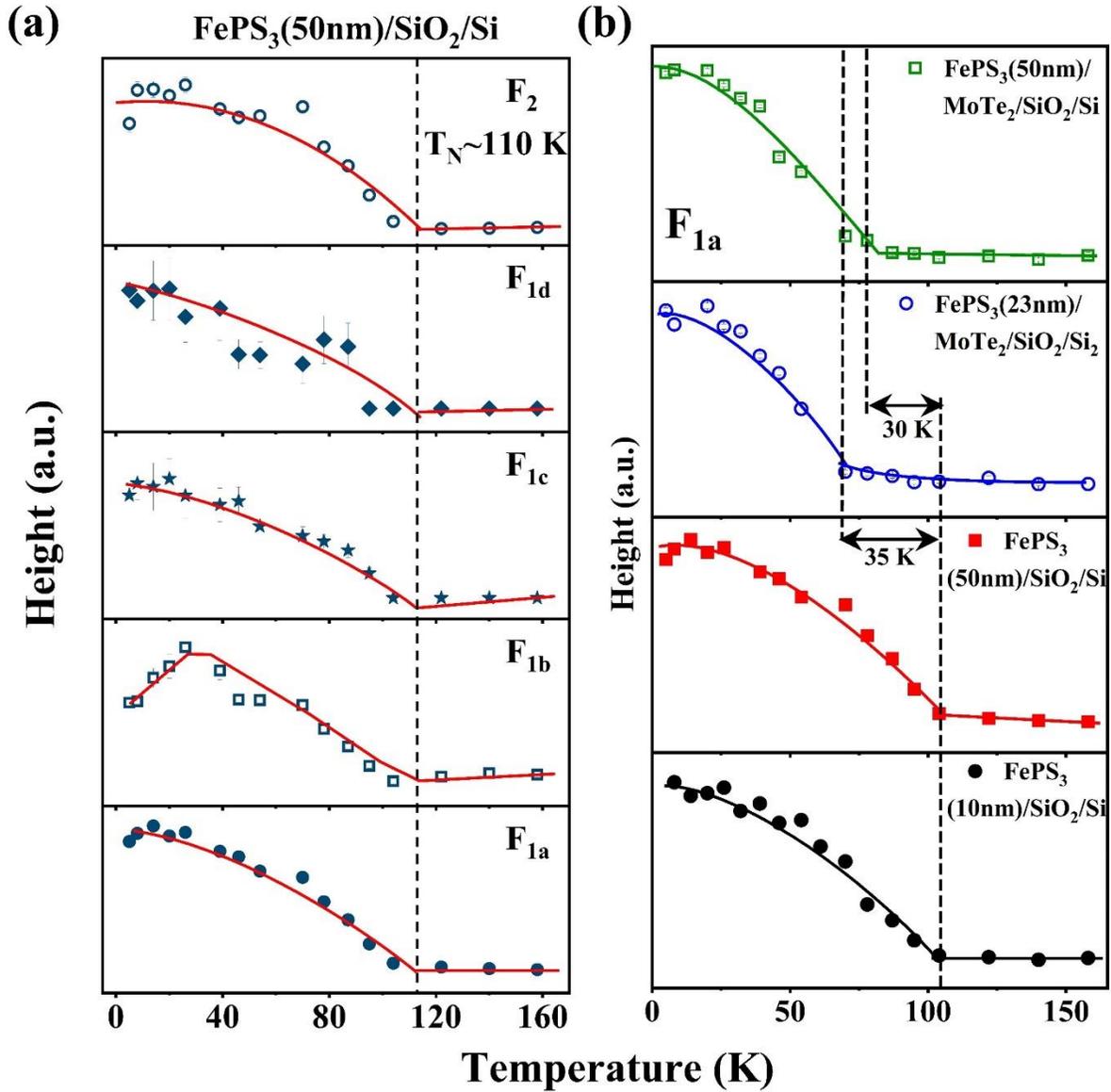

**Figure S11.** The detection of PM to AFM phase transition in FePS$_3$ through Raman measurements. **(a)** The temperature evolution of the modes F$_1$ and F$_2$ of an FePS$_3$ flake (50 nm thick) on SiO$_2$/Si substrate showing clear signatures of the phase transition. While F$_1$ splits into four peaks F$_{1a}$, F$_{1b}$, F$_{1c}$, and F$_{1d}$, the mode F$_2$ narrows with a simultaneous increase in peak height below the Neel temperature. Consequently, the modes F$_{1a}$, F$_{1b}$, F$_{1c}$, F$_{1d}$, and F$_2$ show a dramatic increase in the peak height below the Neel temperature. **(b)** The temperature-dependent peak height of the F$_{1a}$ phonon for various samples of FePS$_3$ on SiO$_2$/Si substrate and MoTe$_2$ underlayer, showing a clear suppression of the Neel temperature for the flakes placed on the underlayer of MoTe$_2$.

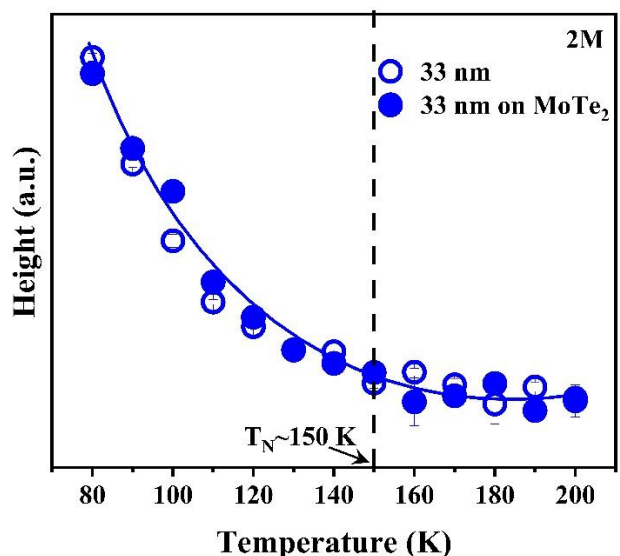

**Figure S12.** The two-magnon (2M) feature exhibits a dramatic decrease in its height above the Neel temperature. The Neel temperatures measured for the 33 nm thick flakes of NiPS3 directly on SiO2/Si substrate (hollow circles) and in the heterostructure with MoTe2 (solid circles) shows no appreciable difference.

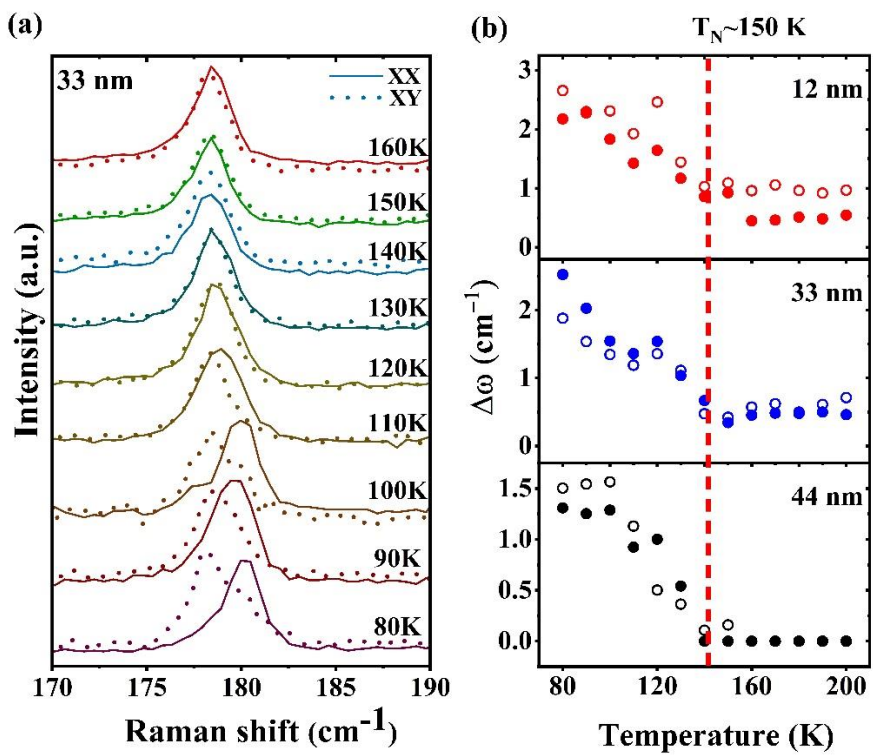

**Figure S13.** (a) The two components of the $N_2$ mode (observed in the xx and xy polarizations, respectively) which become degenerate in the PM phase at various temperatures across the transition. (b) The difference in positions of the two components of $N_2$ ($\Delta\omega$) as a function of temperature, depicting the corresponding Neel temperatures of the various flakes. The flakes of various thicknesses show nearly constant Neel temperature for the flakes directly on SiO2/Si (hollow circles) and in heterostructures with MoTe2 (solid circles).

## SN5. Confirmation of the T$_d$ phase for the MoTe$_2$ underlayers in the heterostructures

The heterostructures of the various MPS$_3$ (M=Mn, Fe, and Ni) 2D antiferromagnets with MoTe$_2$ show suppression of the Neel temperature, which has been extensively discussed in the main text. In this regard, it is important to note that the suppression of the AFM phase has been connected to the Dzyaloshinskii Moriya (DM) interactions, which in turn is associated with strong spin-orbit coupling (SOC) of the non-magnetic layer of the heterostructure (here MoTe$_2$). To ensure the presence of strong SOC, it is important to confirm the phase of the underlying MoTe$_2$ layer in the heterostructure. We must mention that the heterostructures were prepared at room temperature using single crystals of metallic MoTe$_2$ (in the 1T′ phase). However, it is known that the 1T′ phase in bulk MoTe$_2$ undergoes a structural transition to the T$_d$ phase (which is known to be a Type II Weyl semimetal) below ~250 K. The T$_d$ phase is associated with strong SOC which gives rise to the topologically non trivial character to the low temperature phase. In our experiments, the magnetic ordering temperature has been studied for the MnPS$_3$, FePS$_3$, and NiPS$_3$ 2D antiferromagnets, all of which have their T$_N$'s way below ~250 K. Therefore, it is safe to assume that for all the experiments, in the range of investigated temperatures, the MoTe$_2$ underlayers were in the T$_d$ phase. As a further confirmation, we can refer to the Raman spectra of the heterostructures obtained at low temperature (Figure S14). The phase transition from 1T′ to T$_d$ phase can be probed through Raman by the splitting of the P$_6$ mode [33,47] labelled in Figure S14. We observe that for the heterostructures of MnPS$_3$ and FePS$_3$ layers with MoTe$_2$, the P$_6$ mode (of MoTe$_2$) is split into two components, confirming the T$_d$ phase. In case of NiPS$_3$ based heterostructures, due to the proximity of the P$_6$ mode of MoTe$_2$ to the N$_1$ phonon of NiPS$_3$, it is not possible to resolve the P$_6$ mode or its split components at low temperature. Therefore, for the NiPS$_3$ based heterostructures, we have confirmed the T$_d$ phase of MoTe$_2$ by the appearance of the low-frequency sheer mode labelled A which is also reported to appear as a result of the structural transition (to the T$_d$ phase) undergone by MoTe$_2$ [47].

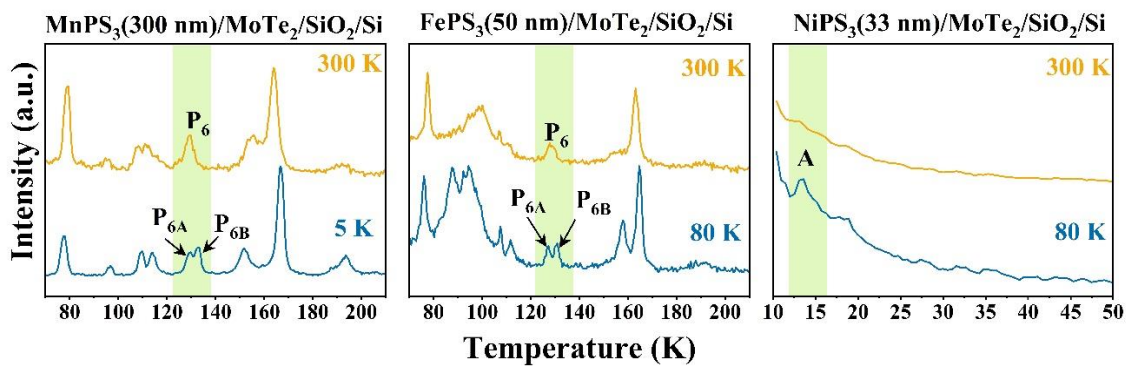

**Figure S14. Confirmation of the T$_d$ phase for the MoTe$_2$ underlayers in the heterostructures.** For the MnPS$_3$ and FePS$_3$ based heterostructures, we clearly observe the splitting of the P$_6$ mode. For NiPS$_3$ based heterostructure, we observe the appearance of the low-frequency sheer mode labelled A. All these signatures confirm that the underlayers of MoTe$_2$ were all in the Td phase in the investigated samples at low temperature.

**SN6. Raman study performed on thin flakes of MnPS$_3$ supported on MoTe$_2$**

The heterostructures of MnPS$_3$ flakes of various thicknesses with MoTe$_2$ have been studied in detail. Unfortunately, the low scattering cross section associated with the thin flakes of MnPS$_3$ results in very weak Raman signal which is difficult to resolve. The Raman signals of these thin flakes can, however, be improved to a great degree when the flakes are supported on SiO$_2$/Si substrate due to the optical interference phenomena. Such an enhancement is not possible when the thin flakes are supported on MoTe$_2$ layers. Therefore, it becomes difficult to resolve the Raman signal for thin layers. The detection of the magnetic ordering phenomena through the evolution of the M$_1$ phonon, therefore, becomes very difficult. In the main text, we have reported the suppression of AFM ordering in MnPS$_3$ flakes (of 144 nm and 300 nm thicknesses) by the influence of the underlayer of MoTe$_2$. We observe that the T$_N$ for the MnPS$_3$ flake of 144 nm thickness is already suppressed to temperatures as low as 5 K or below. Therefore, it is intuitive that any flakes of thinner out-of-plane dimension would show a stronger suppression of the magnetic ordering. Figure S15 shows the weak presence of the M$_1$ mode in MnPS$_3$(14nm)/MoTe$_2$/SiO$_2$/Si sample. At 5 K, the M$_1$ mode appears at ~155 cm$^{-1}$, thereby showing negligible shift with respect to the PM phase signature. This confirms that the AFM ordering in the MnPS$_3$(14nm)/MoTe$_2$/SiO$_2$/Si sample occurs below 5 K (if at all).

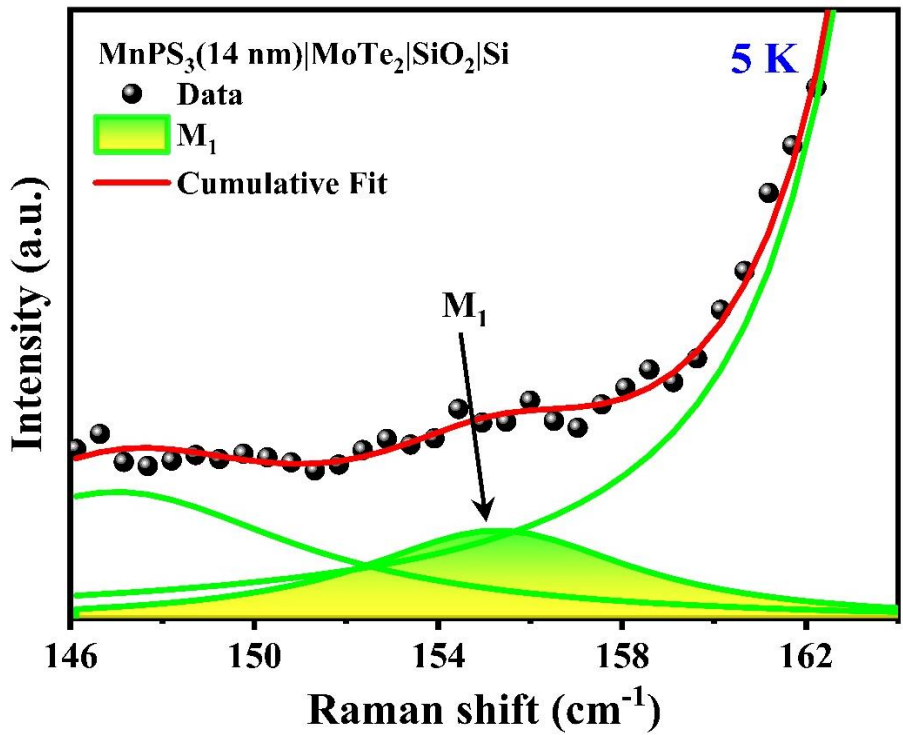

**Figure S15.** The deconvolution of the Raman signal of MnPS$_3$ (14 nm) /MoTe$_2$/SiO$_2$/Si using Lorentzian multi-function clearly showing the M$_1$ phonon of MnPS$_3$ at ~155 cm$^{-1}$, confirming the paramagnetic phase.

## SN7. Robustness of the AFM phase to flake thickness

We observe that the $T_N$'s of the various flakes of $MnPS_3$, $FePS_3$, and $NiPS_3$, supported on $SiO_2$/Si substrate, show no appreciable variation with flake thickness. This is consistent with various recent experimental and theoretical reports [21–23,30,31]. In this regard, it may be useful to note that the Mermin–Wagner–Hohenberg theorem [48,49] predicts that thermal fluctuations on isotropic 2D magnets result in gapless long-wavelength spin wave excitations with finite density of states, thereby preventing a long-range 2D magnetic ordering. However, the presence of an anisotropy (e.g., easy-axis, easy-plane) can lead to the violation of the Mermin–Wagner–Hohenberg theorem by opening a gap in the spin wave spectrum, and allow the formation of magnetic ground states in the 2D magnets at finite temperatures [14]. Such anisotropies can result from the breaking of inversion symmetry, strong spin-orbit interactions, magnetic dipole-dipole interactions, etc. The persistence of magnetism in a layered magnetic material down to atomically thin flakes, therefore, depends on the possibility of the presence of an anisotropy in the 2D limit as the thickness is reduced. In case of $MPS_3$ compounds, the crystal possesses an intrinsic anisotropy in the crystal as a consequence of the trigonal distortion (of the $MS_6$ octahedra) and spin-orbit coupling, resulting in a Hamiltonian as follows [13]:

$$H = -2\sum\{J_\perp(S_{ix}S_{jx} + S_{iy}S_{jy}) + J_\parallel(S_{ix}S_{jx} + S_{iy}S_{jy}) + AS_{iz}^2\} + H_{DM} \quad \ldots (1)$$

where $J_\perp$ and $J_\parallel$ represent the perpendicular and parallel exchange interactions between the spins $\boldsymbol{S_i}$ and $\boldsymbol{S_j}$, while $A$ represents the single-ion anisotropy induced by the axial distortion and spin-orbit coupling. The last term $H_{DM}$ requires the breaking of inversion symmetry along with strong spin-orbit coupling (SOC) and can be safely ignored for flakes of $MPS_3$ supported on $SiO_2$/Si. Therefore, we may expect the single-ion anisotropy term to dominate for these samples. It may however be noted that while the interactions governing the magnetic ordering in $FePS_3$ and $NiPS_3$ have been attributed to the single ion anisotropy [50], $MnPS_3$ shows negligible single-ion anisotropy and its magnetic ordering is predominantly dictated by dipolar interactions [13]. On account of the weak interlayer coupling in these materials, the trigonal distortion and the resultant anisotropy does not show any appreciable change as a function of the flake-thickness. Therefore, the insensitivity of the spin ordering temperature to the flake thickness has been attributed to the weak interlayer coupling exhibited by these materials. Vaclavkova *et al.* [31] reported the effective exchange interaction energies in $MnPS_3$ as 1.11 meV (for intralayer) and 0.04 meV for interlayer nearest neighbors. Kim and Park [30] calculated the interlayer and intralayer exchange interactions in $MnPS_3$, $FePS_3$, and $NiPS_3$ and reported that the interlayer exchange interactions are much weaker than the intralayer interaction strengths in $MnPS_3$ and $FePS_3$. However, for $NiPS_3$, they reported significantly large interlayer interactions (which are comparable to the intralayer exchange interactions), which is likely to result in some thickness dependence of the $T_N$ [30]. However, such dependencies were predicted for the ultrathin flakes of monolayer or bilayer thickness [30]. In the flakes of $NiPS_3$ we studied, ranging from 12 to 44 nm of thickness, we did not observe any thickness dependence for the $T_N$, consistent with previous experimental reports [22].

**SN8. Effect of strain and charge transfer on the AFM phase of the magnetic layer**

Our extensive experiments on heterostructures of MPS$_3$ compounds with T$_d$-MoTe$_2$ reveal a strong suppression of the AFM transition in MnPS$_3$ and FePS$_3$ but, in contrast, no effect on the AFM transition of NiPS$_3$. Our observations are further supported by experiments on the heterostructures with a variety of substrates/underlayers such as topological insulator Sb$_2$Te$_3$, as well as metallic Au and Cu. Our experiments on MnPS$_3$ further show the suppression of the T$_N$ to be stronger for thinner flakes, thereby introducing a thickness-dependent magnetic ordering. In order to look for the possible origin for the above findings, we have enlisted the potential contributors to the observed suppression of the AFM ordering, which are: (i) effect of strain, (ii) effect of charge transfer from the substrate, (iii) SOC associated with the non-magnetic underlayer, and (iv) the spin orientation associated with the magnetic layer. The effect of SOC and the spin orientations have been discussed in the context of the appearance of DM interactions. The other possibilities discussed in detail below.

**Effect of strain:** The AFM ground state of the MPS$_3$ compounds may be altered by means of application of strain, as predicted by Chittari *et al.* [35] using *ab initio* calculations. Therefore, we first take up strain as being the potential reason for the suppression of the AFM ground state as observed in our experiments. Here, we may note that the effect of strain on thin flakes of 2D materials is customarily characterized by the shifts in their phonon frequencies. The application of a tensile (compressive) strain is expected to show a redshift (blueshift) of the phonon frequencies with respect to the unstrained case. In our measurements, we observe a negligible shift in the phonon frequencies of the MnPS$_3$ heterostructures (with T$_d$-MoTe$_2$) with respect to the MnPS$_3$ flakes supported on SiO$_2$/Si substrate (Fig. S16). Importantly, we do not find any noticeable shift in frequencies with varying thickness. This implies that the effect of strain, if at all, is either absent or similar on all the flakes of MnPS$_3$, irrespective of their thicknesses. Hence, the thickness dependence observed in the suppression of the AFM state is not due to strain. Further, as shown in Fig. S17, we have performed experiments on MnPS$_3$ flakes of varying thicknesses by using a variety of substrates/underlayers, including SiO$_2$/Si, T$_d$-MoTe$_2$, Sb$_2$Te$_3$, Au, and Cu. We observe a suppression of the AFM phase with all the substrates/underlayers except SiO$_2$/Si. Out of the various substrates/underlayers used, some (*i.e.*, T$_d$-MoTe$_2$ and Sb$_2$Te$_3$) are single crystals with different dimensions of crystallographic axes while others are amorphous. It is highly unlikely that such a range of varying substrates/underlayers would have the same or comparable strain effects on the magnetic layer. Instead, we may notice that the suppression of the AFM phase is always associated with a substrate/underlayer that has a high charge carrier density and/or spin-orbit coupling. This, thus, rules out strain as the origin for the suppression of the magnetic ordering and leads us to the next possibility, that of a charge transfer from the substrate to the magnetic layer.

**Effect of charge transfer:** Chittari *et al.* [35] have further predicted a transition of the MPS$_3$ systems from the semiconducting AFM state to a metallic FM phase by means of charge doping. It must be noted that the above-mentioned phase transitions are predicted for a charge doping of the order of ~$10^{14}$/cm$^2$. For a geometry (heterostructures) like ours where the 2D magnetic layer lies on a conductive substrate/underlayer without any applied electrical excitation, (e.g., bias or gating), such a high concentration of charge carriers (doping) is unlikely to be available in the high bandgap semiconductors (*i.e.*, MPS$_3$'s), merely through diffusion process across the interface. Therefore, we can safely rule out charge doping as the potential contributor to the suppression of the AFM phase.

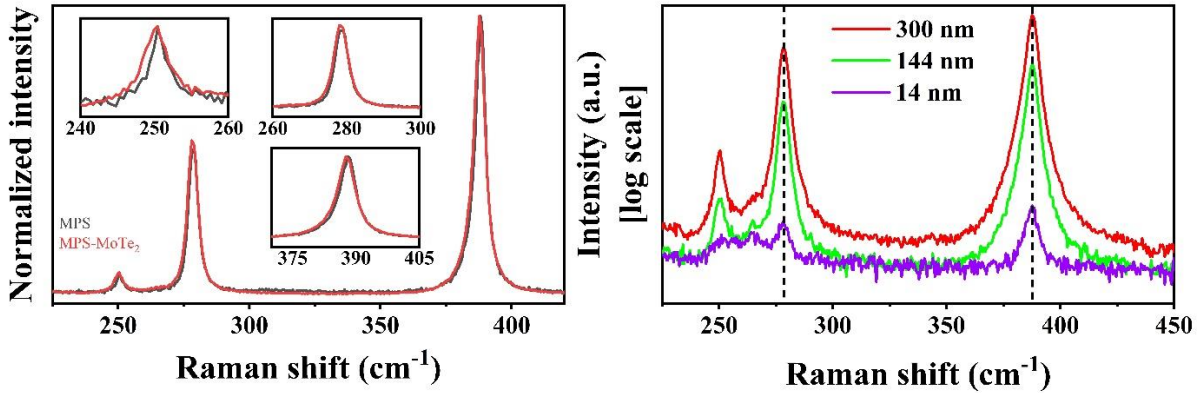

**Figure S16. Effect of strain.** Left panel: The room temperature spectra corresponding to a heterostructure of MnPS$_3$ with MoTe$_2$ shows redshifts in phonon frequencies of all the modes with respect to an MnPS$_3$ flake without the influence of the MoTe$_2$ underlayer. This means the MnPS$_3$ layers in the heterostructure are possibly under the influence of slight tensile strain. Right panel: The spectra for the various heterostructures (with varying thickness of the MnPS$_3$ layer) are compared. We observe no shift in the phonon frequencies as the thickness of MnPS$_3$ is varied in the heterostructures, implying that the strain on MnPS$_3$ due to MoTe$_2$ is independent of thickness.

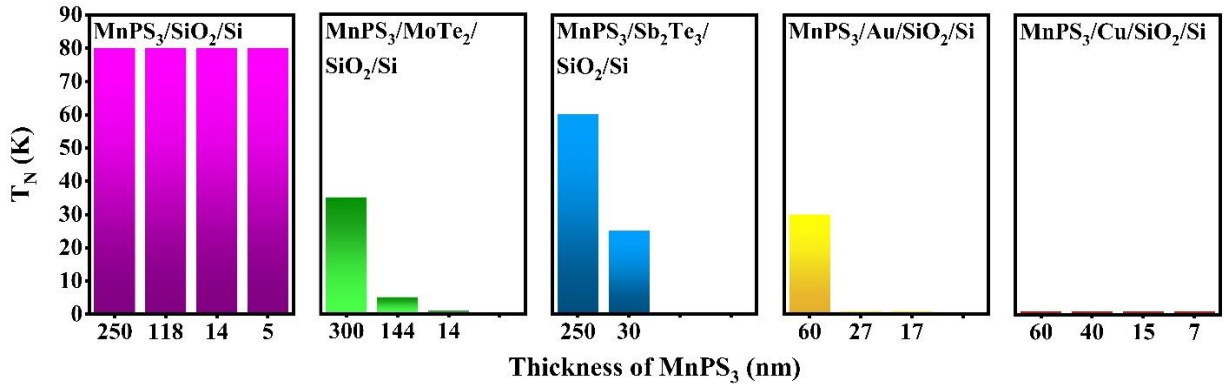

**Figure S17.** The Neel temperatures (T$_N$'s) corresponding to the various flakes of MnPS$_3$ of varying thicknesses transferred on to various substrates/underlayers.

## SN9. Intensity of the Raman modes in the heterostructures

The intensity of the Raman signal is influenced by several factors including the polarizability of the molecules, the intensity and the excitation energy of the excitation laser, the symmetries of the phonons, etc. When a van-der Waals heterostructure is prepared, we expect to observe the signatures of both the constituent layers in the Raman signal obtained from the heterostructure. Further, various types of interactions between the two constituent layers of the heterostructure may cause enhancements or decrements of the Raman signal. In the main text, we have observed suppression in the magnetic ordering temperature of the magnetic layer in the heterostructures with $MoTe_2$ for $MnPS_3$ and $FePS_3$. In this section, we will discuss about certain interesting observations in $FePS_3$ and $MnPS_3$ based heterostructures in relation to their Raman signal intensities.

In the $FePS_3$ heterostructures with $MoTe_2$, we observed a relative increase in the Raman signal intensity of $FePS_3$ induced by the exposure to the underlying $MoTe_2$ (Figure S18). Raman enhancement observed in materials on metallic substrates is caused by an electromagnetic mechanism mediated through surface plasmon resonance phenomena.

In the $MnPS_3$ heterostructures, we observed an interesting trend in the Raman signals of the $MoTe_2$ underlayer. The $MnPS_3$ flakes, acting as capping layers, are expected to optically shield the $MoTe_2$ underlayer appreciably, resulting in a decrease in the effective Raman signal of $MoTe_2$. Further, the signal of $MoTe_2$ is expected to weaken more for heterostructures constituting a thicker capping layer of $MnPS_3$. However, contrary to the expectation, we observe an enhancement in the overall Raman signature of $MoTe_2$ on adding a capping layer of $MnPS_3$. The enhancement is also observed to be stronger for the heterostructures constituting thicker capping layers of $MnPS_3$ (Figure S19). Previously, the use of graphene, h-BN, and $MoS_2$, has been reported to show strong enhancement of Raman signatures of probe molecule [51] through dipolar interactions. The Raman enhancements (of $MoTe_2$) testify to a strong interplanar coupling between the constituent layers of the heterostructures, which further ensures the role of interfacial interactions leading to the tailoring of the magnetic ground states.

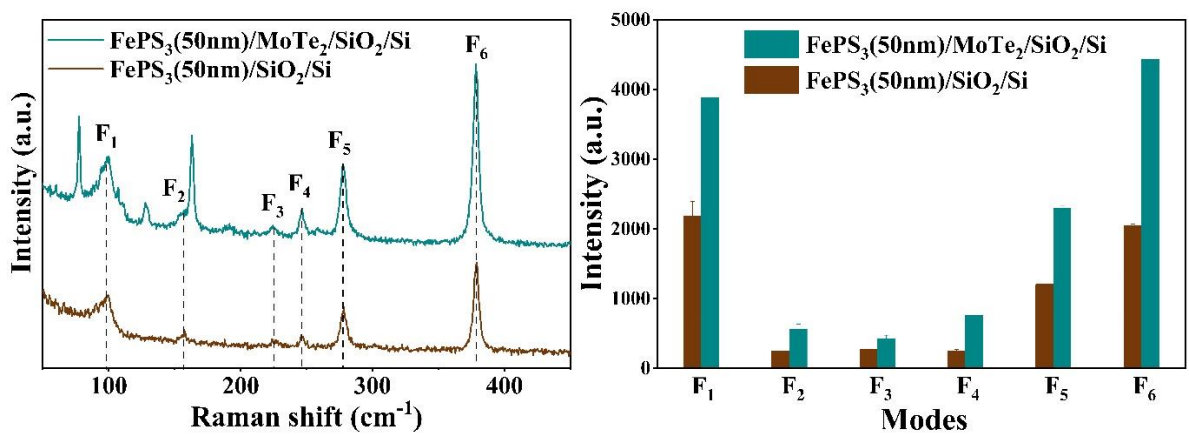

Figure S18. The room temperature Raman spectra of FePS$_3$ compared with the heterostructure of FePS$_3$ with MoTe$_2$. The modes of FePS$_3$ show an enhancement in intensity on introducing the MoTe$_2$ underlayer.

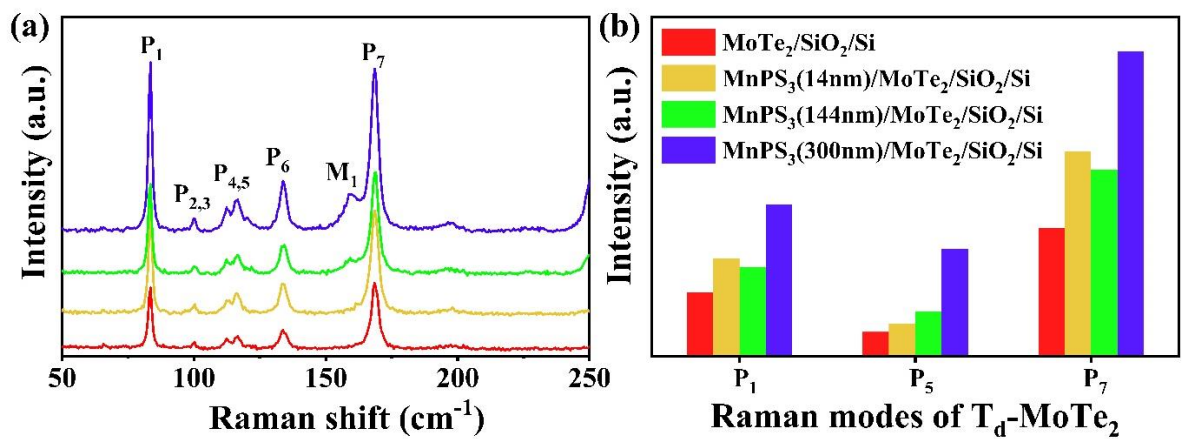

Figure S19. The room temperature Raman spectra of various flakes of MnPS$_3$ with MoTe$_2$ underlayer. The modes of MoTe$_2$ show an enhancement in intensity with increasing thickness of MnPS$_3$.